\documentclass{aa}  
\usepackage{graphicx}
\usepackage{rotating}
\usepackage{txfonts}
\usepackage{hyperref}
\usepackage{placeins}
\begin{document}

\title{The multi-faceted variability of HD~192639:}
\subtitle{Stochastic behaviour, regularities, and an exceptional mass-ejection event\thanks{Based on optical spectra collected with the TIGRE telescope (La Luz, Mexico) and on TESS photometry.}}
\author{Gregor\ Rauw\inst{1} \and Ya\"el\ Naz\'e\inst{1}\thanks{Senior research associate FRS-FNRS (Belgium).} \and Charles-Antoine\ Gilon\inst{1}}
\institute{Space sciences, Technologies and Astrophysics Research (STAR) Institute, Universit\'e de Li\`ege, All\'ee du 6 Ao\^ut, 19c, B\^at B5c, 4000 Li\`ege, Belgium \\ \email{g.rauw@uliege.be}}
\date{}

  \abstract
      {}{Spectroscopic and photometric variability is widespread among O-type supergiants. It is linked to various phenomena affecting the star and its  circumstellar environment, thereby providing direct information concerning them. To investigate such connections, we decided to revisit the prototypical O7.5\,Iabf supergiant HD~192639.}{High-cadence spectroscopic monitoring was performed simultaneously with intensive space-borne photometric observations. The data were analysed with several methods to characterise the variability.}{Besides the usual stochastic, low-frequency photometric variability, our observations reveal the presence of recurrent --but transient-- modulations on a timescale of about five days. The same signal is present in the spectroscopic data and was already seen two decades ago. This stability suggests that this timescale corresponds to the stellar rotation. Furthermore, our observations unveil, for the first time, an unusually strong dimming event in the light curve associated with absorption and emission changes in H\,{\sc i} and He\,{\sc i} lines. This unprecedented trough corresponds to an episodic ejection of a rather large amount of mass (its column density being comparable to that of the steady wind). While rare, such an event could hint at an overlooked aspect of mass loss in massive stars.}{}
\maketitle
\section{Introduction \label{intro}}
Spectroscopic variability is a widespread property of massive early-type OB stars, especially among O supergiants \citep{Ful96}. In presumably single OB stars, low-level stochastic variability of emission lines formed in the stellar wind can arise from the fluctuation of the number of clumps in the wind \citep{Eve98}. More deterministic variations that produce periodic modulations of the morphology of spectral lines can arise from co-rotating large-scale structures inside the stellar winds \citep[e.g.][]{How93,Cra96,Lob08}. Such large-scale structures are usually attributed to evolving brightness inhomogeneities on the stellar surface \citep[e.g.][]{Sud16,Tahina,Rau21}, which could stem from localised magnetic spots \citep{Can11}. Current generations of spectropolarimetric instruments lack the sensitivity to detect such magnetic spots for OB stars \citep{Koc13}. However, these spots alter the local radiative-wind acceleration and generate coherent large-scale spiral structures in the stellar wind known as co-rotating interaction regions \citep[CIRs;][]{Cra96,Lob08}. Another type of co-rotating large-scale structure is linked to the presence of a global dipolar magnetic field \citep[e.g.][]{Sta96,Naz10}, which confines the stellar winds into a region near the magnetic equator \citep{udD13}. About 7\% of the O-type stars feature a global magnetic field strong enough to impact the dynamics of their stellar wind \citep{Gru17}. The magnetic confinement leads to photometric and spectroscopic variability over a wide range of wavelengths, including X-rays \citep[e.g.][]{Naz16}, the UV \citep[e.g.][]{Mar13}, and the optical domain \citep[e.g.][]{Rau23}. Whether the variability comes from CIRs or magnetically confined winds, the observed periodicity is interpreted as the star's rotational period (or an integer fraction of it). In parallel, pulsations of the stellar surface also produce variability of photospheric absorption lines \citep[e.g.][]{Gie99,Kau06,Rau08}. Non-radial pulsations associated with pressure modes (so-called $\beta$-Cep-type pulsations) occur on timescales of hours, whereas gravity-mode pulsations are slower and fall into the same frequency range as the rotation period \citep[e.g.][and references therein]{God17}.

Space-borne photometry collected by missions such as {\it CoRoT}, {\it Kepler}, {\it BRITE,} and {\it TESS} unveiled that OB stars also display widespread low-level photometric variability. The most emblematic characteristics of this photometric variability are not variations at well-defined discrete frequencies, but rather a stochastic low-frequency variability known as red noise \citep[e.g.][]{Blo11,Bow19,She24}. The power spectrum of red noise consists of a forest of closely spaced frequency peaks with amplitudes that increase towards lower frequencies. The individual peaks that make up the red noise have lifetimes from days to weeks as revealed by time-frequency spectra of photometric time series \citep[e.g.][]{Rau21,Naz21}. Three scenarios have been proposed to interpret these red noise variations. One scenario is that they arise from turbulent motion generated in a thin, sub-photospheric, iron-opacity peak convective layer \citep{Can09,Lec19,Lec21,Sch22}. Alternatively, they could correspond to stochastically excited gravity modes arising in the convective core and propagating through the radiative envelope to the stellar surface \citep[e.g.][]{Aer15,Bow19}. Finally, they could also result from wind inhomogeneities generated by the so-called line de-shadowing instability (LDI) of radiatively driven stellar winds \citep{Krt21}. In this case, the power spectra of the photometric variations are best described by a broken power law \citep{Krt21}. Based on their analysis of the {\it TESS} light curves of 150 O-stars, \citet{She24} concluded that, for O-type stars, internal gravity waves likely play a more important role during the early phases of the star's evolution, whilst the iron convection zone would become dominant as the star evolves. Because they found no well-defined correlation between the amplitude of the red noise and the mass-loss rate of the wind, \citet{She24} suggested that clumps play only a limited role in the photometric variability of these stars.

In this article, we revisit the spectroscopic variability of the O7.5\,Iabf supergiant HD~192639 \citep{Sot11}. This star has frequently been considered as an archetypal mid-O supergiant, and its spectrum was repeatedly analysed by means of non-local thermodynamic equilibrium (nLTE) stellar atmosphere models \citep{Pul06,Bou12,Sur13,Haw21,Gor22}. Variability provides complementary information on the star, helping us to understand and interpret its properties. First hints of spectroscopic variability of HD~192639 were reported by \citet{Man55}, \citet{And79}, and \citet{Und95}. More quantitative analyses were presented by \citet{Ful96}, \citet{Rau98}, and \citet{Rau01}. These investigations unveiled that the absorption components of the He\,{\sc ii} $\lambda$\,4686 and H$\alpha$ line profiles undergo recurrent variations on a timescale of roughly 4.8\,days. \citet{Rau01} interpreted these variations as the result of a rotational modulation of the amount of stellar wind material along the line of sight. This interpretation was questioned by \citet{Mar05}, which suggested instead that the variability was due to fluctuations of the number of clumps in the wind or to transient features. In this context, it should be noted that no strong, large-scale magnetic field was detected in HD~192639 during the MiMeS spectropolarimetric survey \citep{Gru17}.

To clarify the issue of the recurrence of the variations and characterise the spectroscopic variability of an archetypal O-type supergiant, we organised a new spectroscopic monitoring campaign of HD~192639 that was coordinated with {\it TESS} photometric observations. Section\,\ref{sect:obs} presents the observations analysed in our study. Sections\,\ref{varlight} and \ref{varspec} report our analysis of the photometric and spectroscopic variability. Finally, we discuss the origin of the observed variations in Sect.\,\ref{discuss}.  

\section{Observations and data processing \label{sect:obs}}
\subsection{Spectroscopic data \label{obs:spectro}}
Spectroscopic observations of HD~192639 were collected in 2021 with the refurbished HEROS echelle spectrograph \citep{Kaufer2,Schmitt} on the fully robotic 1.2\,m TIGRE telescope \citep{Schmitt,Gon22} at La Luz Observatory near Guanajuato (Mexico). The journal of the observations is given in Table\,\ref{Journal}. Each spectrum had an exposure time of 30\,minutes. Our campaign started in April 2021, with typically one observation every other night for one month. A much more intensive spectroscopic monitoring (with up to 14 spectra per night separated by half an hour) took place between 20 July and 16 August 2021. This intensive campaign largely overlapped with the {\it TESS} photometric monitoring of Sector\,41 (see Sect.\,\ref{photom}). The TIGRE/HEROS spectra have a spectral resolving power of 20\,000 over the optical range from 3760 -- 8700\,\AA,\ with a small gap around 5600\,\AA. These spectra were reduced with the HEROS reduction pipeline \citep{Mittag,Schmitt} and were further processed using {\sc iraf} and {\sc midas}. Telluric absorption lines around the He\,{\sc i} $\lambda$\,5876 and H$\alpha$ lines were removed by means of the {\tt telluric} tool within {\sc iraf} using the atlas of telluric lines of \citet{Hinkle}. All spectra were continuum-normalised using {\sc midas} adopting best-fit spline functions adjusted to the same set of continuum windows. 

\subsection{Photometry \label{photom}}
High-cadence, high-precision photometry of HD~192639 was obtained at seven epochs with the Transiting Exoplanet Survey Satellite \citep[{\it TESS},][]{TESS}. {\it TESS} collects broadband photometry over the 6000\,\AA\, to 1\,$\mu$m passband. {\it TESS} monitors sectors covering $24^{\circ} \times 96^{\circ}$ for a nearly continuous duration of about 27\,days. HD~192639 was observed in Sectors 14 \& 15 (18 July -- 11 September 2019), 41 (23 July -- 20 August 2021), 54 \& 55 (9 July -- 1$^{\rm st}$ September 2022), 75 (30 January -- 26 February 2024), and 81 (15 July -- 10 August 2024). Except for the first two sectors, HD~192639 was a target star for which fully processed 2\,min cadence light curves are available on the Mikulski Archive for Space Telescopes (MAST) portal.\footnote{http://mast.stsci.edu/} These light curves, processed with the {\it TESS} pipeline \citep{Jen16}, provide simple background-corrected aperture photometry (SAP) as well as so-called pre-search data-conditioned (PDC) photometry obtained after correcting trends correlated with systematic spacecraft or instrumental effects. We only kept data points with a quality flag of 0 and converted the fluxes into magnitudes. The formal photometric accuracies are $\sim 0.2$\,mmag. Since the PDC data appear to be the most accurate, we focused our analysis on these data. For Sectors 14 and 15, 30-min-cadence light curves were extracted from the full-frame images (FFIs). The FFIs were processed with the Python software package Lightkurve. Aperture photometry was extracted on $51 \times 51$ pixels image cutouts. For the source mask, we adopted a ﬂux threshold of 30 times the median absolute deviation over the median ﬂux. The background was evaluated from those pixels of the image cutouts that were below the median ﬂux, and background subtraction was performed by means of a principal component analysis including five components.\footnote{A simple median evaluation of the background yields similar or slightly lower quality light curves.}

Since the {\it TESS} CCD detectors have pixel sizes of (21\arcsec )$^2$ on the sky and since the aperture photometry is extracted over several pixels, we checked the {\it GAIA} data release 3 catalogue \citep[DR3,][]{DR3} for sources within a 1\arcmin\ radius of our target. {\it GAIA}-DR3 lists 210 such sources, but they are all much fainter than HD~192639, with the two brightest neighbouring sources being 6.5\,mag fainter than our star. Contamination of the {\it TESS} photometry by neighbouring sources is thus negligible, as also confirmed by the value ($\sim 0.98$) of the CROWDSAP keyword.

Table\,\ref{journalTESS} provides some information on the {\it TESS} sectors used for our study. The standard deviations of the photometric data exceed the mean photometric errors by a large factor. This demonstrates that HD~192639 exhibits genuine photometric variability.  

\begin{table*}
  \caption{{\it TESS} sectors used in our analysis. \label{journalTESS}}
  \begin{center}
  \begin{tabular}{c c c c c c c}
    \hline
    Sector & Dates         & $\Delta$\,t & n & Mean error & $\sigma$ \\
           & (HJD-2450000) &    (s)      &   &     (mmag) & (mmag) \\
    \hline
    14 \& 15 &  8683.37 --  8737.39 & 1800 &  2378 & 0.053 &  8.64 \\
    41       &  9419.99 --  9446.58 &  120 & 18315 & 0.215 & 13.09 \\
    54 \& 55 &  9769.90 --  9824.27 &  120 & 36777 & 0.216 &  9.30 \\
    75       & 10339.78 -- 10367.48 &  120 & 18957 & 0.219 &  9.62 \\
    81       & 10506.56 -- 10533.18 &  120 & 18406 & 0.214 &  8.99 \\
    \hline
  \end{tabular}
  \end{center}
  \tablefoot{For each sector, column 2 provides the total time span, whilst columns 3 and 4 indicate the time step and the number of data points. Column 5 quotes the mean photometric error on an individual data point, whereas column 6 provides the standard deviation of the photometric data.}
\end{table*}

\section{Photometric variability \label{varlight}}
\begin{figure}
  \begin{center}
    \resizebox{8.7cm}{!}{\includegraphics{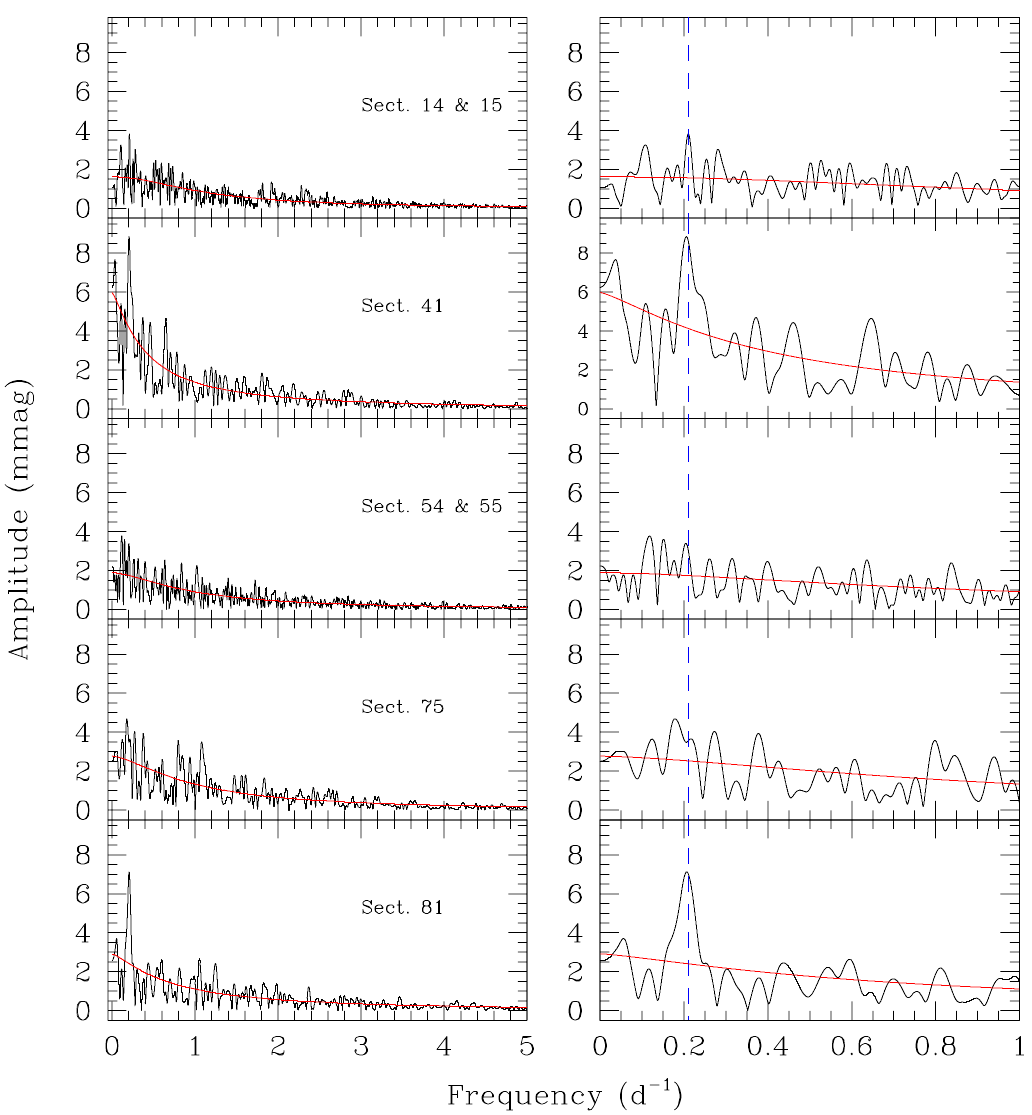}}
  \end{center}  
  \caption{Fourier periodogram of {\it TESS} photometry of HD~192639 during the different sectors. The left-hand panels display the periodogram between 0 and 5\,d$^{-1}$ clearly unveiling the red-noise behaviour. The best-fit red-noise relation is shown in red. The right-hand panels provide a zoomed-in view of the region below 1\,d$^{-1}$. The dashed blue vertical line yields the $\nu_1$ frequency previously found in spectroscopic time series by \citet{Rau01}. \label{sp_montage}}
\end{figure}

The {\it TESS} data allowed us to search for photometric variability on timescales from under ten minutes and up to two months in the case of consecutive sectors. The light curves unveil modulations at the few mmag level on timescales of a few days. However, there does not appear to be a single fully stable recurrence time. To characterise these variations, we analysed the {\it TESS} data with the Fourier method of \citet{HMM} amended by \citet{Gos01}. The epoch-dependent periodograms are displayed in Fig.\,\ref{sp_montage}. One can clearly see an increase of the amplitude towards lower frequencies, which is typical of the red noise that is commonly seen in space-borne photometry of many categories of massive stars \citep[e.g.][]{Blo11,Bow19,Bow20,Naz20,Rau21,Naz21,Naz24,She24}.

We used the formalism of \citet{Sta22} to adjust the red noise in the periodograms. We thus fitted an expression,
\begin{equation}
  A_{\rm rn}(\nu) = \frac{A_0}{1 + (2\,\pi\,\tau\,\nu)^{\gamma}} + C_{\rm wn}
,\end{equation}
with $A_{\rm rn}(\nu)$ being the red noise contribution to the amplitude (in mmag) at frequency $\nu$ in the periodogram. The parameters $A_0$ (scaling factor), $\gamma$ (slope), $\tau$ (mean lifetime), and $C_{\rm wn}$ (white noise level) were determined by means of a Levenberg–Marquardt algorithm. The errors on the parameters were estimated from the diagonal elements of the variance–covariance matrix with a renormalisation of the $\chi^2$ \citep[for details, see][]{Naz21}. Since there are no strongly outstanding discrete peaks in the periodograms, we assumed here that the periodograms are solely due to stochastic low-frequency variations, and we thus fitted the red-noise relation without any prior pre-whitening of the periodograms. The best red-noise relations are shown by the red curves in Fig.\,\ref{sp_montage}, and their parameters are listed in Table\,\ref{fitparam}.

\begin{table*}
  \caption{Best-fit red-noise parameters and properties of the highest peak in the periodograms of the {\it TESS} light curves. \label{fitparam}}
  \begin{center}
  \begin{tabular}{c c c c c c c}
    \hline
    Sector & $A_0$ (mmag) & $\tau$ (d) & $\gamma$ & $C_{\rm wn}$ (mmag) & $\nu_{\rm peak}$ (d$^{-1}$) & $A(\nu_{\rm peak})/A_{\rm rn}(\nu_{\rm peak})$ \\
    \hline
    14 \& 15 & $1.62 \pm 0.05$ & $0.140 \pm 0.006$ & $1.90 \pm 0.10$ & $0.016 \pm 0.007$ & 0.2103 & 2.3 \\
    41 & $5.74 \pm 0.06$ & $0.371 \pm 0.007$ & $1.41 \pm 0.02$ & $0.002 \pm 0.001$ & 0.2057 & 2.1 \\
    54 \& 55 & $1.85 \pm 0.01$ & $0.161 \pm 0.002$ & $1.67 \pm 0.02$ & $0.004 \pm 0.001$ & 0.1180 & 2.1 \\    
    75 & $2.68 \pm 0.03$ & $0.158 \pm 0.003$ & $1.69 \pm 0.02$ & $0.005 \pm 0.001$ & 0.1786 & 1.8 \\
    81 & $2.81 \pm 0.04$ & $0.215 \pm 0.005$ & $1.44 \pm 0.02$ & $0.006 \pm 0.001$ & 0.2065 & 2.9 \\
    \hline
  \end{tabular}
  \end{center}
\end{table*}

Apart from the red noise, there are no clearly outstanding permanent peaks. It is remarkable, however, that the strongest peaks (in Sectors 41 and 81) are precisely located at the $\nu_1 = 0.21$\,d$^{-1}$ frequency that was previously found in spectroscopic variations by \citet[][see Fig. \ref{sp_montage}]{Rau01}. The highest peaks in the periodograms of Sectors 14 \& 15 also correspond to that frequency, although with even lower significance. 

\begin{figure}
  \begin{center}
    \resizebox{8.7cm}{!}{\includegraphics{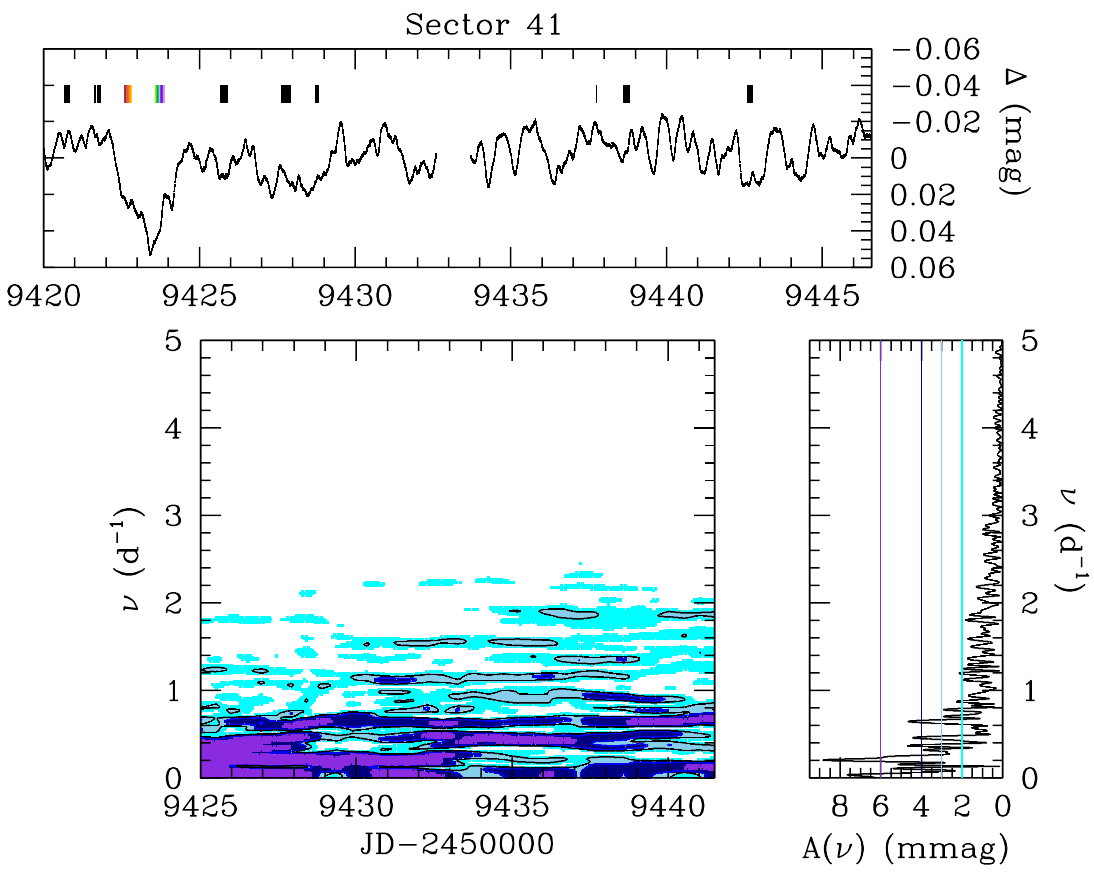}}
  \end{center}
  \caption{Time–frequency diagram of Sector 41 {\it TESS} data of HD~192639. The top panel displays the {\it TESS} light curve. The bottom left-hand panel illustrates the evolution of the Fourier periodogram, with the date corresponding to the middle of the 10\,d sliding window. The violet, dark blue, light blue, and cyan colours, respectively, represent areas with amplitudes $\geq 6$, $\geq 4$, $\geq 3,$ and $\geq 2$\,mmag. The bottom right panel provides the Fourier periodogram evaluated over the full light curve of Sector 41. The various coloured lines correspond to the colour-scale used in the bottom left panel. The tick marks in the top panel indicate the times of our TIGRE observations. Those in colours correspond to the trough event. \label{Sect41}}
\end{figure}

Besides computing periodograms for each sector (or combinations of consecutive sectors), we also built time-frequency diagrams. For this purpose, we computed Fourier periodograms of the photometric data extracted over a sliding temporal window with a 10\,d duration and shifted in steps of 1\,d. As an illustration, the time-frequency diagram of Sector 41 is presented in Fig.\,\ref{Sect41}. Others are shown in Figs. \ref{Sect1415} -- \ref{spevol5}. This allows us to more easily detect transient features. For example, in Sector 14, the strongest peak in the periodogram is found at a frequency of $0.208 \pm 0.004$\,d$^{-1}$ (period of 4.76\,d), while it shifted to $0.268 \pm 0.004$\,d$^{-1}$ (period of 3.73\,d) during Sector 15. Such transient features are seen in the time-frequency diagrams of other massive stars. For instance, the Onfp stars\footnote{Also known as Oef stars.}  where spectroscopic variability is suggested to arise from short-lived, magnetically rooted localised prominences \citep{Sud16} display transient timescales in their spectroscopic and photometric variability \citep{Rau21}.

Another interesting feature is worth highlighting. A few days after the beginning of Sector 41 observations, around JD\,245923.5, the star's brightness dropped by $\sim$0.05\,mag (hereafter the `trough event'; see Fig.\,\ref{Sect41}). While it is not the sole example of a brightness drop, it is the most extreme event of its kind in terms of depth and duration among the seven {\it TESS} light curves of HD~192639 (duration of $\sim 3$\,d with a drop of 0.05\,mag vs. duration $< 1.5$\,d with drops $< 0.04$\,mag; see Figs. \ref{Sect1415} -- \ref{spevol5}). To the best of our knowledge, such events have not been reported before in the light curves of O-type stars. Fortunately, our spectroscopic monitoring was scheduled to largely overlap with Sector 41, and the spectroscopic data enable us to gain further insight into the nature of this special event (see Sect.\,\ref{dis_trough}).

The best-fit parameters of the red noise model for Sector 41 differ quite significantly from those found for the other sectors. Indeed, the periodogram has a rather strong peak at 0.037\,d$^{-1}$, that is the inverse of the duration of the sector, thus indicating a long-term trend. We tested whether this is due to the trough event by repeating the periodogram calculation after excluding data taken before HJD\,2459425. However, the amplitude at the lowest frequencies was only slightly lowered (to 5.7\,mmag), and the $A_0$ scaling factor remained high ($3.77 \pm 0.05$\,mmag). Therefore, the trough event alone cannot explain the difference in red noise properties during Sector 41.

\section{Spectroscopic variability\label{varspec}}
\begin{table*}
  \caption{Strongest peaks found in Fourier periodograms of RV data. \label{RVfreq}}
  \begin{center}
  \begin{tabular}{c l}
    \hline
    Line & \multicolumn{1}{c}{Strongest peaks in periodogram} \\
    \hline
    He\,{\sc ii} $\lambda$\,4200 & 0.259\,d$^{-1}$ (3.8\,km\,s$^{-1}$); 0.268\,d$^{-1}$ (3.9\,km\,s$^{-1}$); 0.698\,d$^{-1}$ (3.9\,km\,s$^{-1}$) \\ 
    H$\gamma$ & 0.167\,d$^{-1}$ (7.2\,km\,s$^{-1}$); 0.555\,d$^{-1}$ (7.4\,km\,s$^{-1}$); 0.845\,d$^{-1}$ (7.0\,km\,s$^{-1}$) \\     
    He\,{\sc i} $\lambda$\,4471 & 0.554\,d$^{-1}$ (4.7\,km\,s$^{-1}$) \\
    He\,{\sc ii} $\lambda$\,4542 & 0.268\,d$^{-1}$ (3.3\,km\,s$^{-1}$) \\
    N\,{\sc iii} $\lambda$\,4634 & 0.136\,d$^{-1}$ (6.5\,km\,s$^{-1}$) \\
    N\,{\sc iii} $\lambda$\,4640 & 0.207\,d$^{-1}$ (8.3\,km\,s$^{-1}$); 0.402\,d$^{-1}$ (8.2\,km\,s$^{-1}$)\\
    H$\beta$ & 0.147\,d$^{-1}$ (16.0\,km\,s$^{-1}$); 0.156\,d$^{-1}$ (16.2\,km\,s$^{-1}$); 0.166\,d$^{-1}$ (16.3\,km\,s$^{-1}$); 0.176\,d$^{-1}$ (16.2\,km\,s$^{-1}$) \\     He\,{\sc ii} $\lambda$\,5412 & 0.305\,d$^{-1}$ (4.5\,km\,s$^{-1}$); 0.539\,d$^{-1}$ (4.6\,km\,s$^{-1}$); 0.548\,d$^{-1}$ (4.6\,km\,s$^{-1}$); 0.697\,d$^{-1}$ (4.7\,km\,s$^{-1}$) \\
    \hline
  \end{tabular}
  \tablefoot{The numbers in parentheses indicate the associated amplitude.}
  \end{center}
\end{table*}

\subsection{Radial velocities \label{radvel}}
Radial velocities (RVs) were measured on the normalised spectra by adjusting Gaussians to the line profiles of strong absorption lines. The $\sigma_{RV}$ of the interstellar Ca\,{\sc ii} K absorption line amounts to 0.5\,km\,s$^{-1}$. This value can be seen as an indicator of the intrinsic RV accuracy of the wavelength calibration of our data. The RVs of the stellar lines have dispersions that significantly exceed this accuracy. These large differences cannot be explained by the larger width of the stellar lines compared to the interstellar absorptions. The smallest dispersions for stellar lines were obtained for the He\,{\sc ii}$\lambda\lambda$\,4200, 4542, and 5412 lines, which have $\sigma_{RV} = 5.4$, $4.1$ and $5.6$\,km\,s$^{-1}$, respectively. Since these lines form deep within the photosphere of the star, they should provide the most stringent constraints on genuine stellar motion. The largest dispersions were obtained for the H\,{\sc i} Balmer lines with $\sigma_{RV} = 8.6$ and $17.7$\,km\,s$^{-1}$ for H$\gamma$ and H$\beta$, respectively. The latter lines are heavily contaminated by circumstellar emissions.  

\begin{figure}
  \begin{center}
    \resizebox{8.7cm}{!}{\includegraphics{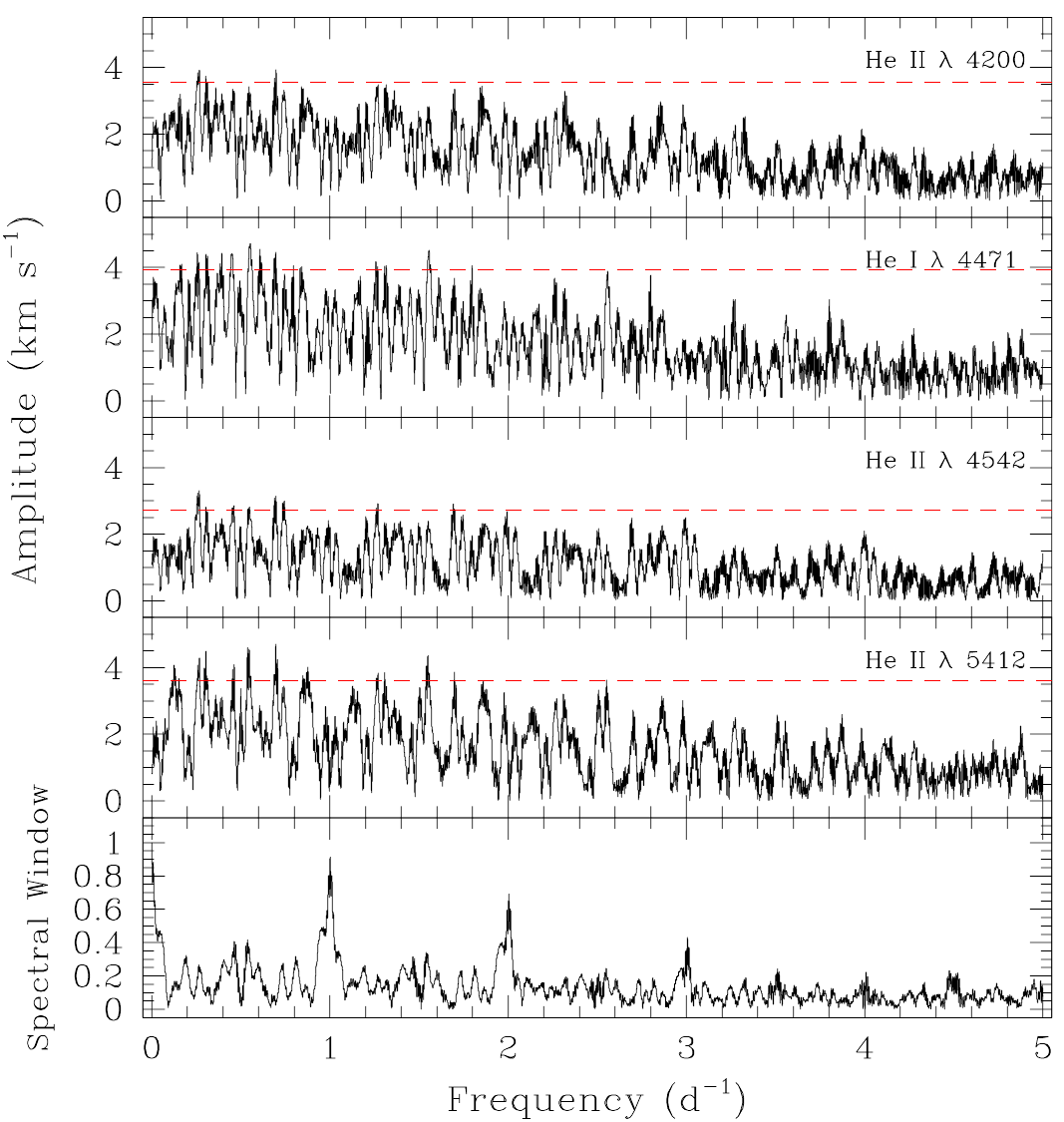}}
  \end{center}  
  \caption{Top four panels: Fourier periodograms of RV time series of He lines in the optical spectrum of HD~192639. The dashed red line yields the 1\% significance level (see text). The bottom panel shows the spectral window corresponding to our time series. \label{RVperiod}}
\end{figure}

The RV time series were analysed with the same Fourier periodogram method \citep{HMM,Gos01} as applied to the photometric data in Sect.\,\ref{varlight}. This method explicitly accounts for the gaps in ground-based astronomical time series. As an illustration, Fig.\,\ref{RVperiod} displays the periodograms for He\,{\sc i} and He\,{\sc ii} lines along with the spectral window of the time series. These plots further display a 1\% significance level for each periodogram. These significance levels were obtained via a bootstrapping method, where the 108 pairs of times and RVs were mixed randomly, and the reshuffled artificial time series was analysed with the same Fourier method. This process was repeated 1000 times, and for each realisation of the time series the amplitude of the highest peak in the Fourier periodogram was recorded. From the histogram of these amplitudes, we then determined the threshold value, which is such that 99\% of the amplitudes of the reshuffling simulations fall below it.

As one can see, the periodograms do not display any clearly dominant peak. In general, the amplitude increases towards lower frequencies, similarly to the red noise behaviour as seen in photometry, although the aliasing blurs the picture. Those peaks in the periodogram with the highest amplitude are only slightly above the 1\% significance threshold. Table\,\ref{RVfreq} lists the strongest peaks found in the Fourier periodograms of the RV values of the individual lines. As appears clearly from inspection of Fig.\,\ref{RVperiod} and Table\,\ref{RVfreq}, different lines yield different peaks. No frequency is consistently found in the periodogram of each of the lines. The only frequencies that are common to several lines concern the He\,{\sc ii} ion, where peaks near frequencies of 0.26 - 0.27 and 0.70\,d$^{-1}$ are present in the Fourier periodograms of all three lines. Our analysis thus does not reveal any significant and consistent periodicity that could be attributed to orbital motion. This result agrees with the conclusions of the study of \citet{Rau01}, which inferred an upper limit of 5\,km\,s$^{-1}$ on the amplitude of RV variations of He\,{\sc ii} $\lambda$\,4542 and found no consistent periodicity in RV data. Based on seven observations of the He\,{\sc i} $\lambda$\,5876 line, \citet{Kob22} classified HD~192639 as a multiple star candidate. From our data, it appears that this line can display strong profile variations that are not due to orbital motion, nor to pulsations, but reflect the intrinsic variability of the supergiant (see Sect.\,\ref{specvar}). 

\begin{figure}
  \begin{center}
    \resizebox{8.7cm}{!}{\includegraphics{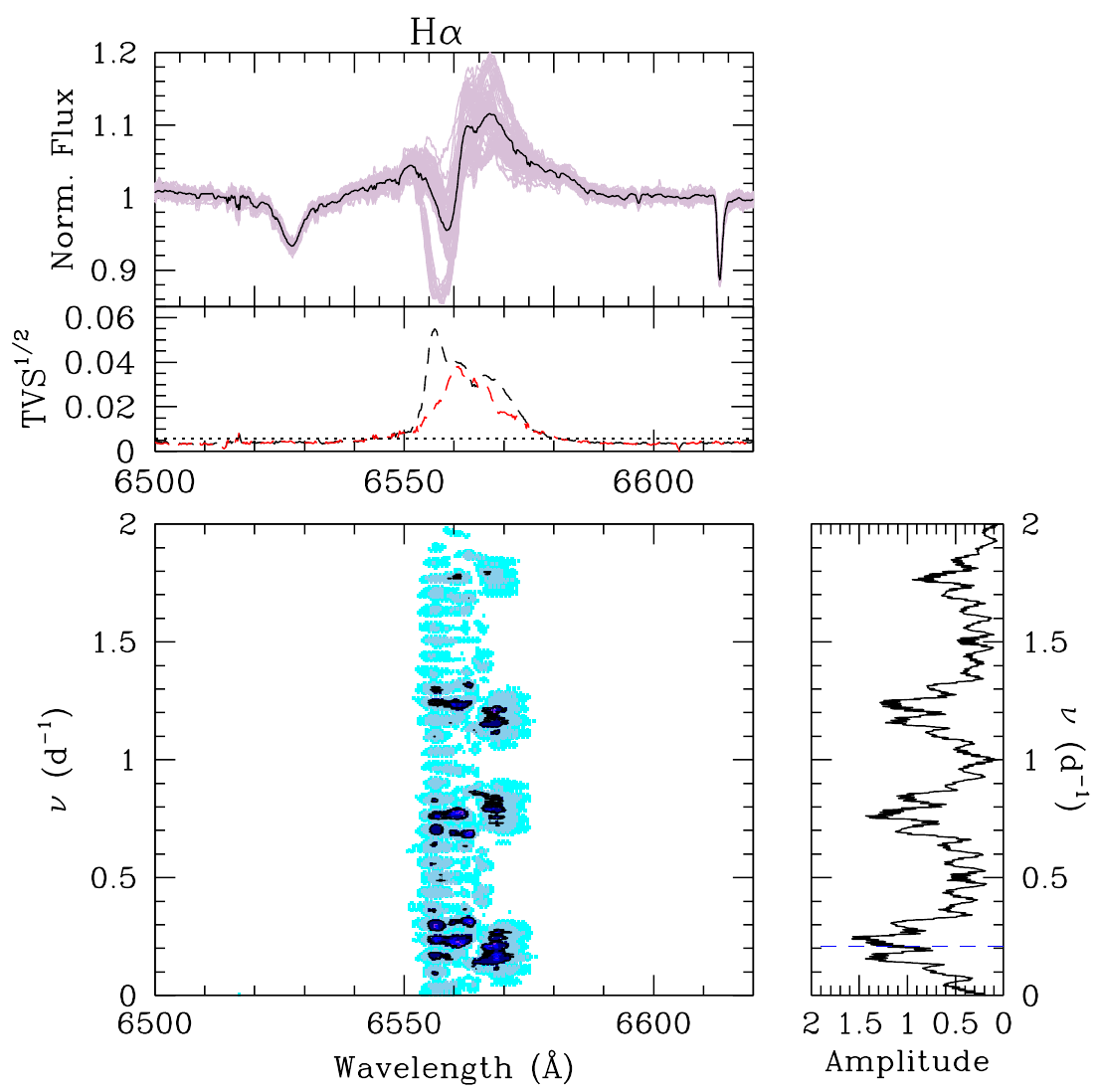}}
  \end{center}
  \caption{TVS and Fourier analysis of our time series of H$\alpha$ line profiles. The top panel illustrates the mean profile on top of all the individual profiles (grey) as well as the TVS$^{1/2}$ (dashed line). The dotted line corresponds to the 1\% significance level assessed from the S/N of the data in the continuum. A TVS$^{1/2}$ level above this threshold has a less than 1\% probability of being due to a noise fluctuation. The dashed red line corresponds to the TVS$^{1/2}$ evaluated excluding the observations taken during the nights affected by the trough event (from HJD\,2459422.6 to 2459425.9). The bottom left panel provides the Fourier periodogram as a function of wavelength divided by the 1\% significance level. The colours indicate ratios of 0.5 (cyan), 1.0 (medium blue), 2.0 (dark blue), and 3.0 (violet). The periodogram divided by the 1\% significance level averaged over the wavelength interval where significant variability occurs is shown by the rightmost panel. The dashed blue line corresponds to $\nu_1$. \label{TVS_Ha}}
\end{figure}

\begin{figure*}
 \begin{minipage}{8.7cm}
  \begin{center}
    \resizebox{8.7cm}{!}{\includegraphics{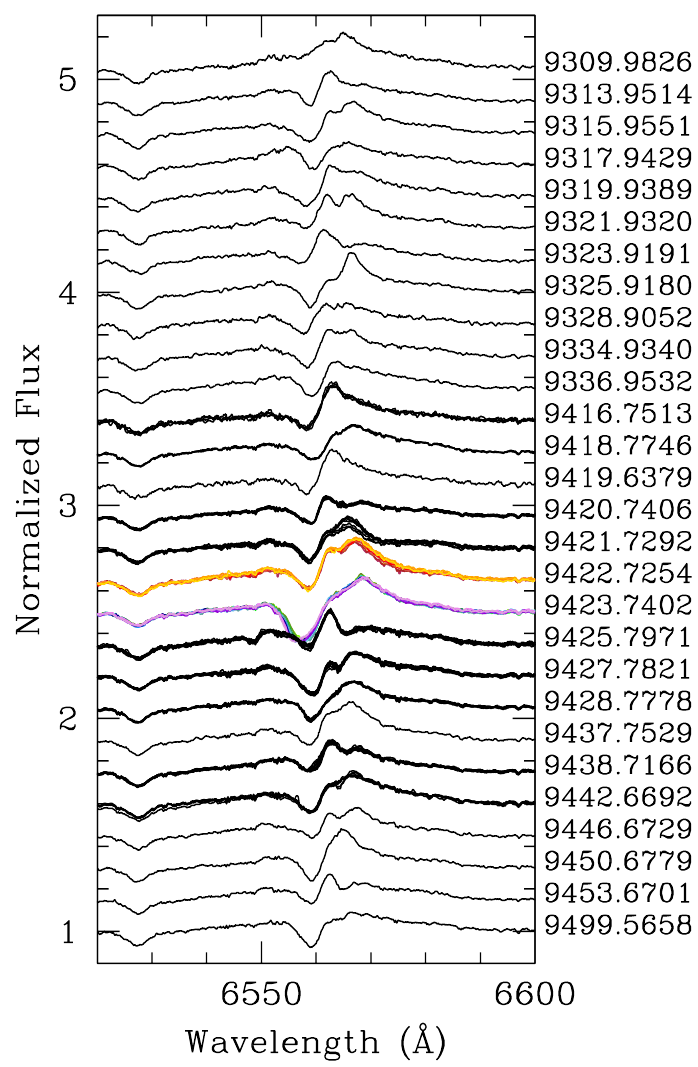}}
  \end{center}
 \end{minipage}
 \hfill
 \begin{minipage}{8.7cm}
  \begin{center}
    \resizebox{8.7cm}{!}{\includegraphics{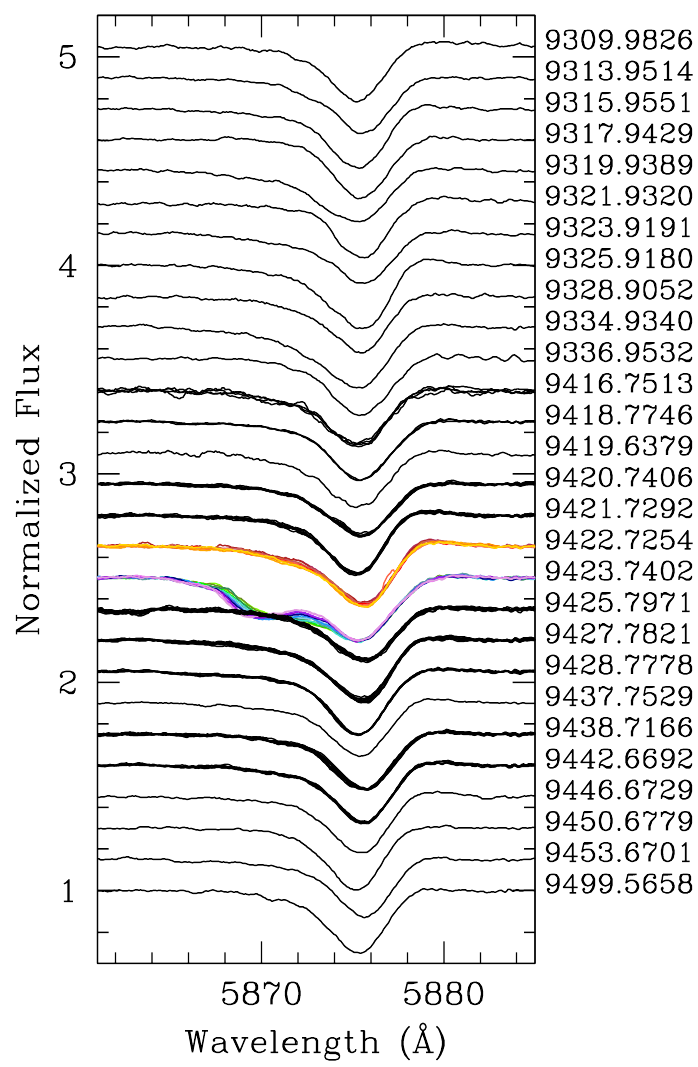}}
  \end{center}
 \end{minipage}
  \caption{Illustration of lpv of H$\alpha$ (left panel) and He\,{\sc i} $\lambda$\,5876 (right panel) lines in our TIGRE/HEROS spectra of HD~192639. The date of the observation is given on the right of each panel in the HJD-2450000 format. For nights with more than one observation, the various spectra are overplotted and the date corresponds to the mean time at mid exposure of all the spectra taken that night. The spectra displayed in orange-red and green-blue colours were taken during the trough seen in the {\it TESS} light curve of Sector\,41.\label{montage}}
\end{figure*}

\subsection{Projected rotational velocity}
Literature values of the projected rotational velocity, $v\,\sin{i}$, of HD~192639 span a rather wide range between 80\,km\,s$^{-1}$ \citep{Mar15} and 125\,km\,s$^{-1}$ \citep{Her92}. Part of this range could reflect the impact of macroturbulence. Over the last two decades, most $v\,\sin{i}$ determinations relied either on the Fourier transform method --which is in principle insensitive to macroturbulence-- or on a goodness of fit approach explicitly accounting for the presence of macroturbulence \citep{Sim14,Sim17,Hol22}. In the case of HD~192639, \citet{Sim14} and \citet{Sim17} applied these methods to the O\,{\sc iii} $\lambda$\,5592 line and derived $v\,\sin{i} = 98$ -- $104$\,km\,s$^{-1}$ with a macroturbulence of $96$ -- $98$\,km\,s$^{-1}$. Using the same line and the same methods, but applied this time to spectra from nine different epochs, \citet{Hol22} inferred a significantly lower projected rotational velocity of $v\,\sin{i} = 82$\,km\,s$^{-1}$ with a macroturbulence of $95$\,km\,s$^{-1}$.

We applied the Fourier transform method to the O\,{\sc iii} $\lambda$\,5592 and C\,{\sc iv} $\lambda$\,5812 line profiles extracted from the mean spectrum of the TIGRE/HEROS data (see Fig.\,\ref{vsini}). The best-fitting rotational broadening functions have $v\,\sin{i} = (110 \pm 5)$\,km\,s$^{-1}$ for the O\,{\sc iii} line and $v\,\sin{i} = (115 \pm 5)$\,km\,s$^{-1}$ for the C\,{\sc iv} line. These fits are of limited quality, however. This could be due, at least partially, to some variability in the C\,{\sc iv} lines \citep{Ful96}.

\subsection{Line-profile variability \label{specvar}}
As previously described in \citet{Rau98} and \citet{Rau01}, some lines in the spectrum of HD~192639 display prominent line-profile variations (lpv). To quantitatively assess the level of lpv in the TIGRE spectra, we computed the temporal variance spectrum \citep[TVS,][]{Ful96} of our spectroscopic time series. 

The photospheric absorption lines He\,{\sc ii} $\lambda\lambda$\,4200, 4542, and 5412 do not display any significant lpv, whereas significant variations --well above the 1\% level-- affect the He\,{\sc i} $\lambda\lambda$ 4471, 5876, 6678, C\,{\sc iii} $\lambda\lambda$\,4647-50-51, He\,{\sc ii} $\lambda$\,4686, H$\beta,$ and H$\alpha$ lines (Figs.\,\ref{TVS_Ha} and \ref{TVS_bis}). Figure\,\ref{montage} shows the variations of the H$\alpha$ profile. The depth of the absorption at $\sim 6560$\,\AA\ and the extent of its blue wing vary. The morphology of its emission component also strongly changes: it appears most of the time double-peaked, but with considerable variations in the relative strength of the peaks. Changes of a similar nature are seen in the other Balmer lines. The variations of the He\,{\sc i} absorption mostly concern the blue wing of the line, whilst the red wing remains essentially constant (see Fig. \ref{montage} for the case of He\,{\sc i} $\lambda$\,5876). The variations of the He\,{\sc ii} $\lambda$\,4686 line follow the same behaviour as previously described in \citet{Rau98} and \citet{Rau01}: the line is nearly filled up with emission, and the profile changes from a weak single emission to apparent P-Cygni or apparent inverted P-Cygni or double-peaked emission profiles. Finally, the C\,{\sc iii} $\lambda\lambda$\,4647-50-51 triplet undergoes a global modulation of its strength, alternating between weak but definite absorptions and a weak broad emission bump. 

\citet{Mar05} argued that the spectroscopic variability of HD~192639 could be explained by random fluctuations in the distribution of clumps across the wind. Such a process should lead to an asymmetric single-peaked TVS with a reduced variability over the red emission component due to cancellation effects of the fluctuations in the emission lobes \citep{Har00,Mar05}. Instead, a CIR would result in a double-peaked, symmetrical TVS profile if the full period of rotation were sampled (\citealt{Har00}; \citealt{Mar05}; see also \citealt{Mart15} for an attempt to quantify the connection between the TVS and different types of variability). Neither of these descriptions matches the variety of TVS behaviours that we observe for different lines. Whilst the TVS of the He\,{\sc ii} $\lambda$\,4686 line displays a double-peaked morphology, this is not the case for the other lines. We notably draw attention to Fig.\,\ref{TVS_Ha}, which clearly demonstrates that the shape of the TVS strongly depends on the sampling of the variability by the data. The same conclusion can be reached by comparing Figs.\,\ref{TVS_Ha} and \ref{TVS_bis} of this work with the TVS illustrations shown in \citet{Rau01} and \citet{Mar05}. Thus, the shape of the TVS alone does not allow us to draw conclusions about the origin of the variability. 

We applied our Fourier periodogram method as a function of wavelength to those lines that show significant variability. To assess the significance of the different peaks in the periodogram, we used the same reshuffling process as in Sect.\,\ref{radvel}, applied to the time series extracted at eight different wavelengths across the H$\beta$ and H$\alpha$ lines. The chosen wavelength sampled different levels of the TVS and allowed us to establish a linear scaling relation between the value of $TVS^{1/2}$ and the 1\% significance level. We then used these scaling relations to assess the significance of the detected peaks (see results in the bottom panels of Figs.\,\ref{TVS_Ha} and \ref{TVS_bis}). 

In the case of the H$\alpha$ line, the periodogram is dominated by a group of three sub-peaks, with the highest sub-peak at 0.241\,d$^{-1}$ (exceeding the 1\% significance level by a factor of more than 3), along with their aliases. The triple structure of this feature cannot be attributed to aliasing. Indeed, Fig.\,\ref{RVperiod} shows no such triple features in the spectral window of the spectroscopic time series. The most likely explanation for these features is that they trace a phenomenon with a frequency close to $\nu_1$, but that was only present with a stable period and phase for a limited duration.

A very similar situation holds for the H$\beta$ and He\,{\sc i} $\lambda\lambda$\,4471 and 5876 lines, which all exhibit the highest peak in their periodograms at 0.177\,d$^{-1}$. These peaks are in all cases part of a much broader structure roughly centred on 0.2\,d$^{-1}$. In the case of the C\,{\sc iii} $\lambda\lambda$\,4647-50-51 absorptions, we find two peaks, one at 0.240\,d$^{-1}$ and the second highest at 0.178\,d$^{-1}$. In addition, the periodogram of this blend further exhibits the signature of long-term variations. In the periodogram of the He\,{\sc ii} $\lambda$\,4686 line, the strongest peak appears at a frequency of 0.206\,d$^{-1}$, which is possibly associated with a harmonic frequency near 0.4\,d$^{-1}$.

By far the strongest variation of the lines is observed on the night HJD\,2459423.7, which occurs during the trough seen in the {\it TESS} light curve of Sector 41 (orange-red and green-blue lines in Fig.\,\ref{montage}). Not only is the absorption in the H$\alpha$ line unusually strong on that night, but its blue edge gradually migrates towards shorter wavelengths over the night. In the He\,{\sc i} $\lambda$\,5876 absorption, we observe the development of a blueshifted absorption appearing around 5871\,\AA. This feature could easily be mistaken as the spectroscopic signature of a secondary star in a binary system, especially since the same behaviour is seen in other He\,{\sc i} lines (He\,{\sc i} $\lambda\lambda$\,4471 and 6678 for instance). Such a discrete absorption component could explain the multiple star status assigned by \citet{Kob22}. However, the fact that the RVs of the He\,{\sc ii} absorption lines do not move and the simultaneous photometric dimming weaken this interpretation. Unfortunately the subsequent night was lost due to poor weather conditions, but the H$\alpha$ spectra of the next night, HJD\,2459425.8, unveil a broad blueshifted emission bump extending from 6550 to 6557\,\AA, which is not seen on other nights. These spectroscopic changes occurring during the trough actually account for the sharp blue peak in the TVS of the H$\alpha$ profile in Fig.\,\ref{TVS_Ha}. Indeed, when we exclude the nights HJD\,2459422.6 -- 2459425.9 from the analysis, this peak disappears from the TVS (red curve in Fig.\,\ref{TVS_Ha}). 

\section{Discussion \label{discuss}}
As pointed out in Sect.\,\ref{intro}, HD~192639 has been analysed with several nLTE spherical stellar atmosphere models \citep{Pul06,Bou12,Sur13,Haw21,Gor22}. In Table\,\ref{modelparam}, we compile the most relevant stellar and wind parameters derived by different authors. \citet{Pul06} analysed H$\alpha$, infrared, and radio data to investigate the stratification of clumping in the stellar wind. \citet{Bou12} applied the CMFGEN model atmosphere code to far-ultraviolet (FUV), ultraviolet (UV), and optical data accounting for optically thin clumps (microclumping). \cite{Sur13} used the PoWR code with the implementation of optically thick clumps (macroclumping) at the wavelengths of the FUV P\,{\sc v} lines. \citet{Haw21} used the FASTWIND code also accounting for macroclumping. It is quite remarkable that these spherical models often fail to correctly reproduce the shape of the He\,{\sc ii} $\lambda$\,4686 line. The best agreement was obtained by \citet{Bou12} and \citet{Haw21}, who analysed the same optical spectrum, taken at a time when the emission in this line was strong (although still less strong than in the model of these authors). On the other hand, \citet{Sur13} and \citet{Gor22} were unable to reproduce the much weaker and more complex He\,{\sc ii} $\lambda$\,4686 line profile seen in their data. This suggests that the innermost regions of the wind, where the He\,{\sc ii} $\lambda$\,4686 emission forms, are strongly impacted by time-dependent non-spherical structures, in agreement with the results of our variability study. 

\begin{table}
  \caption{Stellar and wind parameters from the literature.\label{modelparam}}
  \begin{center}
  \begin{tabular}{c c c c}
    \hline
    $R_*$ (R$_{\odot}$) & $v_{\infty}$ (km\,s$^{-1}$) & $\dot{M}$ ($10^{-6}$\,M$_{\odot}$\,yr$^{-1}$ & Ref. \\
    \hline
    18.5 & 2150 & $\leq 3.0$ & $[1]$ \\
    $20.4 \pm 1.2$ & $1900 \pm 100$ & $1.20 \pm 0.22$ & $[2]$ \\
     & & $1.26$ & $[3]$ \\
    $20.7 \pm 0.6$ & 2700: & $1.41 \pm 0.16$ & $[4]$ \\
    $19.8$ & $1460 \pm 160$ & $1.65 \pm 0.34$ & $[5]$ \\
    \hline
  \end{tabular}
  \tablefoot{References are $[1]$ = \citet{Pul06}, $[2]$ = \citet{Bou12}, $[3]$ = \citet{Sur13}, $[4]$ = \citet{Haw21}, $[5]$ = \citet{Gor22}. Values with a colon (:) were considered dubious by the authors of the study. \citet{Sur13} adopted $R_*$ and $v_{\infty}$ from \citet{Pul06}.}
  \end{center}
\end{table}

\subsection{Order in chaos}
It is remarkable that the analyses of data taken more than 20 years apart and with completely independent instrumentation yield the same $\sim 5$\,d cycles for the variations of the He\,{\sc ii} $\lambda$\,4686 line, as can be seen from the Fourier periodogram in the top right panel of Fig.\,\ref{TVS_bis}. Moreover, this timescale also appears in the analyses of other spectral lines and, even though its amplitude is low, in the photometric variations (especially Sectors 41 and 81). While individual detections may be noisy and the variability pattern changes with epoch, the repeated detection of the same timescale is undoubtedly significant. 

Based on 7 snapshot H$\alpha$ line profiles spread over 18 months with a minimum separation of one month between two consecutive observations, \citet{Mar05} argued that the spectroscopic variability could be explained by random fluctuations in the distribution of clumps across the wind. However, they could not correctly sample the relevant timescale, and our results reveal a different picture. Indeed, the confirmation of a recurrent timescale and the fact that the variations coherently appear over a substantial part of the line profiles would be hard to explain by random fluctuations of the clump distribution. 

\citet{Ful96} argued that pulsations are likely responsible for much of the observed lpv of O-type stars,\footnote{\citet{Ful96} noted that the location of their sample stars in the Hertzsprung-Russell diagram overlapped with the predicted instability region for strange mode oscillations.} especially for photospheric absorption lines. However, in the case of HD~192639, they pointed out that the variability of the C\,{\sc iv} lines is not of purely photospheric origin. In this context, we stress that the analysis of the photospheric He\,{\sc ii} absorption lines of HD~192639 did not unveil any obvious signature of non-radial pulsations, nor does the light curve analysis reveal their presence. Most of the spectroscopic variability of this star thus arises from its stellar wind.

Combining the radii inferred from the spectroscopic analyses (Table \ref{modelparam}) with the $v\,\sin{i}$ values, we infer upper limits on the rotational period between 13.1 and 7.5\,d, depending on the chosen combination. Using the known stellar parameters (Table\,\ref{modelparam}), we also compute the critical break-up velocity, which yields lower limits on the rotational period ranging between 1.9\,d \citep{Pul06} and 2.9\,d \citep{Bou12}. We thus conclude that any periodicity between 2.9 and 7.5\,d, as is the case of the recurrence timescale found here, appears compatible with a rotational modulation.

\citet{Rau01} proposed that the recurrent lpvs reflected either a CIR or a corotating magnetically confined wind. However, no large-scale strong magnetic field was detected for HD~192639 in the MiMeS survey \citep{Gru17}. This result de facto rules out a magnetically confined wind scenario to explain the $\nu_1$ cycle. Moreover, such a configuration should result in characteristic photometric variations \citep{Mun20} that are not observed in the {\it TESS} light curves of HD~192639. The most plausible scenario thus appears to be the presence of CIRs rotating into and out of our sightline. The latter could arise from small-scale localised magnetic fields generated by the subsurface iron-opacity peak convection zone \citep{Can11}. The epoch-dependence of the modulation is not unexpected for such a scenario. Indeed, the surface spots will appear more or less randomly and will have a limited lifetime.

\subsection{C\,{\sc iii} $\lambda\lambda$\,4647-50-51 variability}
An interesting feature is the fact that the visibility of the C\,{\sc iii} $\lambda\lambda$\,4647-50-51 (2s3s\,$^3$S -- 2s3p\,$^3$P$^0$) triplet alternates between clear absorptions and filled-in features. This is reminiscent of the variations in the spectra of so-called Of?p stars. The Of?p category was originally defined by \citet{Wal72} to designate Of stars displaying strong C\,{\sc iii} $\lambda\lambda$\,4647–50-51 emission lines. These stars were later shown to feature strong dipolar magnetic fields that are inclined with respect to the rotation axis (e.g.\ \citealt{Wad15} and references therein). The C\,{\sc iii} $\lambda\lambda$\,4647–50-51 lines undergo strong variations as the stars rotate, changing from absorptions or near absent lines to strong emissions \citep{How07,Naz10,Rau23}. It thus seems tempting to assume that the variations of the C\,{\sc iii} triplet in the spectrum of HD~192639 could also reflect a similar phenomenon, although less extreme than in the case of the Of?p stars.

The formation mechanisms of the C\,{\sc iii} $\lambda\lambda$\,4647-50-51 triplet were studied by \citet{Mar12}. These authors found that these lines are highly sensitive to extreme UV (EUV) transitions. More specifically, it is the C\,{\sc iii} $\lambda$\,538 triplet that drains the 2s3s\,$^3$S level. This process is enhanced by the interplay with the partially overlapping Fe\,{\sc iv} $\lambda$\,538.057 transition that helps to depopulate this level, thus enhancing the C\,{\sc iii} $\lambda\lambda$\,4647-50-51 emission \citep{Mar12}. Stars with significant winds further see their C\,{\sc iii} formation region shifted to higher velocity regions, with the ensuing Doppler shifts leading to desaturation of the key EUV transitions. This further enhances the draining and thus the strength of the C\,{\sc iii} $\lambda\lambda$\,4647-50-51 emission \citep{Mar12}. Regarding the variability of the line in the Of?p stars, \citet{Mar12} proposed that it reflects the existence of a magnetically confined wind (and thus an enhanced draining by desaturation of the EUV lines) in the region of the magnetic equator.
HD~192639 does not harbour a strong dipolar field, which is needed to produce  large-scale confined winds. However, if the variability of HD~192639 is due to spots resulting from {\em \emph{localised}} magnetic fields, then the variations of the C\,{\sc iii} $\lambda\lambda$\,4647-50-51 triplet could have yet another origin. Indeed, Zeeman splitting could alter the overlap between the C\,{\sc iii} $\lambda$\,538 triplet and the Fe\,{\sc iv} $\lambda$\,538.057 transition, possibly enhancing the draining effect of the 2s3s\,$^3$S level, thus leading to an emission that would fill in the C\,{\sc iii} $\lambda\lambda$\,4647-50-51 triplet.

\subsection{The trough event \label{dis_trough}}
The dimming detected at the beginning of Sector 41 lasted for nearly three days. Our spectra unveil simultaneous changes in He\,{\sc i} and Balmer lines, but not He\,{\sc ii}, suggesting the feature to arise in material that is not too hot ($\lesssim 30$\,kK) and hence not too close to the star. The specroscopic variations first consist of an enhanced redshifted emission as the luminosity decreases and then a strongly enhanced absorption near the light-curve minimum. Furthermore, the blue wing of the enhanced absorption shifts bluewards by 75 -- 100 km\,s$^{-1}$ over the 0.27\,d time span of the spectra taken during that night.

The simultaneous photometric dimming and the bluewards-progressing discrete absorption component suggest that this event stems from the attenuation of photospheric radiation by excess material ejected by the star and moving away from it. Either this material was ejected directly into our line of sight or was ejected elsewhere and then rotated into our sightline. The redshifted emission seen in H$\alpha$ during night HJD\,2459422.7 and the subsequent (weaker) blueshifted emission seen during night HJD\,2459425.7 (see Fig.\,\ref{montage}) are reminiscent of material rotating across our sightline, arguing in favour of the second scenario. Furthermore, the duration of the event is long; that is, more than half the rotation period. This indicates that the material covered a wide solid angle. This can happen either because it was ejected over a wide range of longitudes or because the structure is extended, for instance, in a spiral shape, or both. 

The observed drop in brightness by 0.05\,mag corresponds to an increase in optical depth by $\Delta \tau = 0.046$. Considering that the continuum opacity of the stellar wind in the optical band is dominated by Thomson scattering, this number corresponds to a free electron column density of $6.92\,10^{22}$\,cm$^{-2}$. Assuming a solar composition and a fully ionised wind, this electron column density translates into a column density of 0.134\,g\,cm$^{-2}$. We can compare this to the overall wind column density. Assuming a homogeneous wind with a simple $v(r) = v_0 + (v_{\infty} - v_0)\,\left(1 - \frac{R_*}{r}\right)$ velocity law, the density at distance $r$ is expressed as
\begin{equation}
  \rho(r) = \frac{\dot{M}}{4\,\pi\,r^2\,\left[v_0 + (v_{\infty} - v_0)\,\left(1 - \frac{R_*}{r}\right)\right]}
.\end{equation}
The wind column density along the sightline to the star then becomes
\begin{equation}
  \int_{R_*}^{\infty} \rho(r)\,dr = \frac{\dot{M}}{4\,\pi\,R_*\,v_{\infty}}\,\frac{\ln{\frac{v_{\infty}}{v_0}}}{1 - \frac{v_0}{v_{\infty}}}
,\end{equation}
where $\dot{M}$ is the wind mass-loss rate and $R_*$ the stellar radius. 

\begin{figure}
  \begin{center}
    \resizebox{8.7cm}{!}{\includegraphics{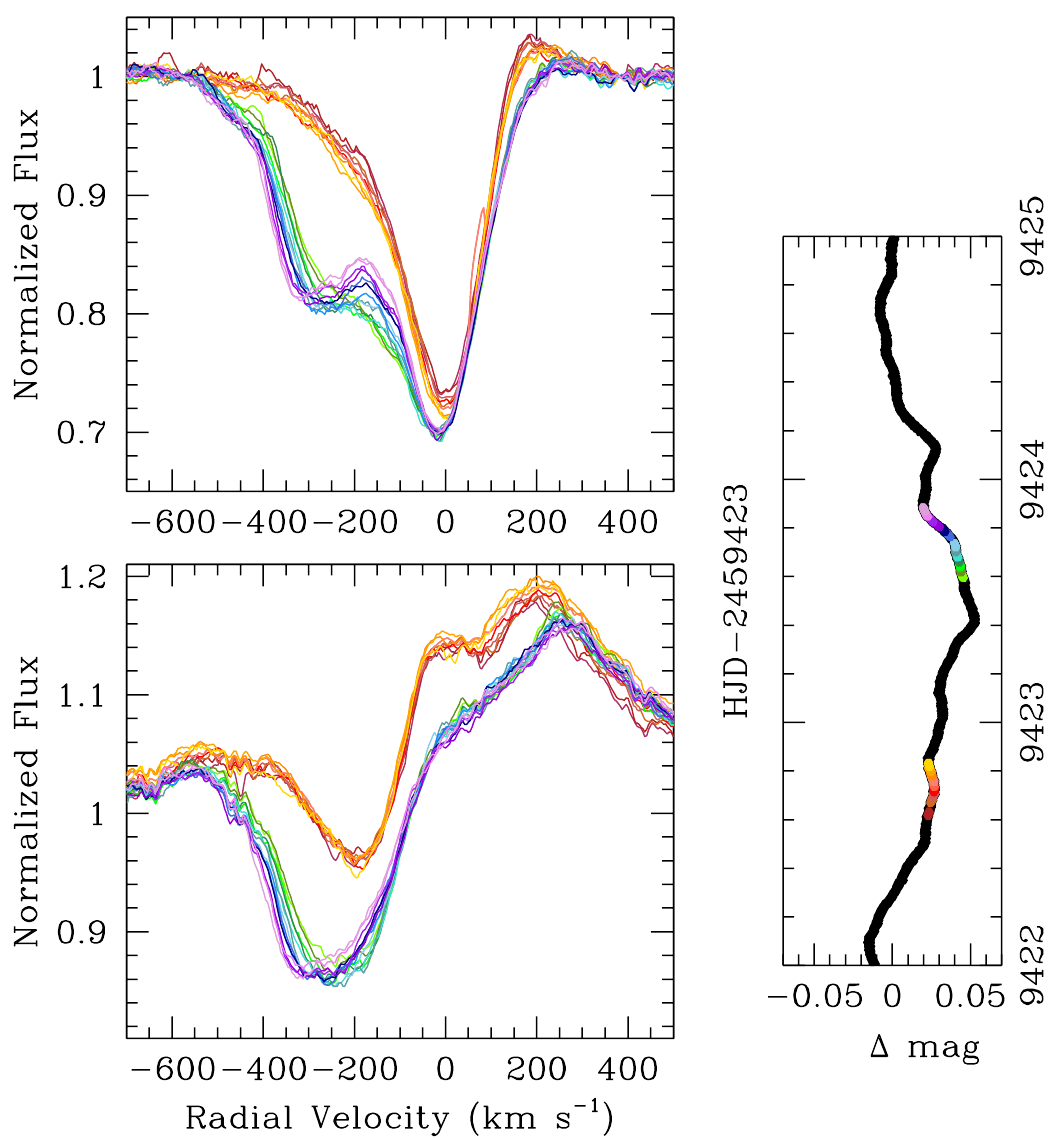}}
  \end{center}
 \caption{Details of lpv during the trough event in {\it TESS} photometry. The top panel illustrates the variations of He\,{\sc i} $\lambda$\,5876, whilst the bottom panel displays those of H$\alpha$. The different colours correspond to different observations taken during the nights of 26 -- 27 and 27 -- 28 July 2021. The vertical panel on the right displays the {\it TESS} light curve. The colours of the symbols correspond to those used for the spectra. \label{DAC}}
\end{figure}

From the parameters in Table\,\ref{modelparam}, we infer values of $\frac{\dot{M}}{4\,\pi\,R_*\,v_{\infty}}$ ranging between 0.0185\,g\,cm$^{-2}$ \citep{Haw21} and 0.0412\,g\,cm$^{-2}$ \citep{Gor22}. If we assume $v_0 = 10$\,km\,s$^{-1}$ at the inner boundary of the wind, the wind column density amounts to 0.104\,g\,cm$^{-2}$ -- 0.207\,g\,cm$^{-2}$. Hence, with a column density of 0.134\,g\,cm$^{-2}$, the trough event corresponds to a density enhancement of about 65 -- 125\% of the nominal wind column density.

The spectroscopic signatures (shifting absorptions) seem reminiscent of the behaviour of discrete absorption components (DACs). These features have been observed to migrate bluewards across the profiles of UV resonance lines of OB stars \citep[e.g.][]{Pri87,Pri98,Ful97,Kap99}. \citet{How89} determined the characteristics of DACs\footnote{\citet{How89} referred to DACs as narrow absorption features.} in a large number of O-type stars observed with the {\it IUE} satellite. They found that the DAC column density typically amounts to 9 -- 30\% of the total column density of the underlying P-Cygni absorption component. This is significantly less than the 65 -- 125\% column-density increase that we inferred for the trough event. A similar conclusion can be reached by comparing the trough column density to the DAC column densities analysed by \citet{Kap99}. The most extreme event studied by these authors was observed for 19~Cep (O9.5\,Ib), reaching a Si$^{3+}$ DAC particle column density of $6.8\,10^{14}$\,cm$^{-2}$. Assuming solar composition and the mean ionisation fraction $\log{\frac{n({\rm Si}^{3+})}{n({\rm Si})}} = -2.88$ \citep{How89}, this translates into a column density of 0.037\,g\,cm$^{-2}$. This value is 3.6 times lower than our estimate for the trough event.    

Whilst DACs are usually detected in UV spectra, the trough is not the first occurrence of a bluewards-migrating absorption excess in the optical domain. A similar feature has also been detected once in the optical spectra of the supergiants HD~152408 \citep[O8:Iafpe,][]{Pri94} and HD~151804 \citep[O8\,Iaf,][]{Pri96}. In the He\,{\sc i} $\lambda$\,5876 absorptions of these stars, the DACs migrate bluewards over a velocity range of $\sim$500\,km\,s$^{-1}$ within $\sim$1\,d. The detection of similar variability in HD~192639, another supergiant O-star, is thus not unexpected. What is quite exceptional is the strength of the absorption feature and the fact that our data unveil, for the first time, the simultaneous broad-band photometric signature of such an event. 

\begin{figure}
  \begin{center}
    \resizebox{8.7cm}{!}{\includegraphics{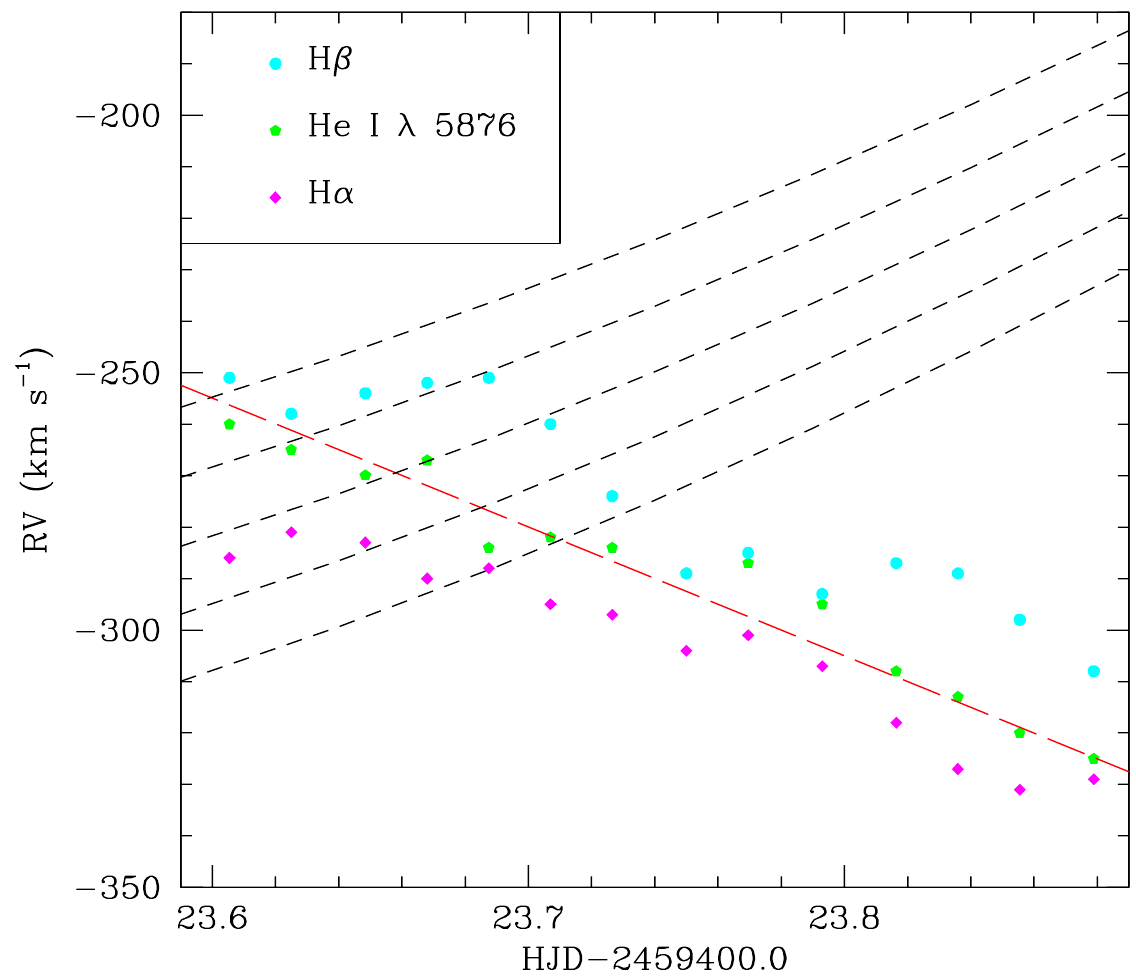}}
  \end{center}
 \caption{RVs of discrete absorption in three spectral lines of HD\,192639 during the rising part of the trough event. Different symbols stand for different lines as shown by the insert. The red long-dashed line corresponds to a bluewards acceleration of 3\,m\,s$^{-2}$. The short-dashed lines correspond to the predictions from Eq.\,\ref{CIRvr} for line-formation regions between 1.17 and 1.21\,R$_*$. \label{DAC_RV}}
\end{figure}

To quantify the motion of the discrete absorption during the night of 27--28 July, we subtracted the mean spectrum of the previous night (26--27 July) from each observation of 27--28 July and fitted a Gaussian absorption to the residuals. The results are illustrated in Fig.\,\ref{DAC_RV} for the H$\beta$, He\,{\sc i} $\lambda$\,5876, and H$\alpha$ lines. On average, we find that the discrete absorption undergoes a $\sim 3$\,m\,s$^{-2}$ bluewards acceleration (see Fig.\,\ref{DAC_RV}), which is quite similar to the $\sim 5.5$\,m\,s$^{-2}$ acceleration reported for HD~152408 \citep{Pri94} and HD~151804 \citep{Pri96}. 

\citet{Ful97} argued that DAC events should be distinguished from more global recurrent modulations of spectral lines (partially) formed in the wind which are attributed to CIRs. Indeed, both phenomena do not necessarily occur on the same timescale. For instance, in the case of HD~64760, \citet{Lob08} found that the DAC recurrence time corresponds to about five times the rotation period, while it occurs with the rotation period or a fraction of it for CIRs. This suggests that the DACs of this star would not be anchored in magnetic spots, but rather result from the interference of three non-radial pulsation modes leading to a retrograde beat pattern \citep{Kau06}. In this context, we recall that neither our spectroscopic monitoring nor the {\it TESS} photometry unveiled signatures of pulsations in HD~192639; hence, we exclude the pulsation-triggered DAC scenario.  

Instead, we decided to test whether a CIR structure could account for the trough event. To this aim, we built a simple model to simulate electron scattering by such a feature. To describe the shape of the CIR, we adopt the semi-analytical description of \citet{Cra96} and \citet{Har00}. In this model, the spiral-shaped CIR is given by
\begin{equation}
  \phi(r) = \phi(R_*) - \frac{v_{\rm rot}}{v_{\infty}}\,\int_1^{r/R_*}\frac{1 - s^{-2}}{w_0 + (1-w_0)(1-s^{-1})}\,ds
,\end{equation}
where $\phi(r)$ is the azimuthal angle at distance $r$; $v_{\rm rot}$ is the equatorial rotational velocity; $v_{\infty}$ is the terminal wind velocity; and $w_0 = \frac{v_0}{v_{\infty}}$, where $v_0$ is the velocity at the inner wind boundary (taken here to be 10\,km\,s$^{-1}$). Following \citet{Har00}, we assumed that the thickness of the CIR in the direction perpendicular to the spiral path varies as $0.1 \times r$ and that the CIR arises from stellar latitudes between $-30^{\circ}$ and $+30^{\circ}$. Hydrodynamic simulations of CIRs \citep{Cra96,Lob08} indicate density contrasts between the CIR and the normal wind, which are typically $\leq 5$. In our default model (see Table\,\ref{CIRparam}), we assumed $\rho_{\rm CIR}/\rho_{\rm wind} = 5.0$. To further maximise the effect of the CIR, we assumed that we observed the star equator-on.

The stellar surface is discretised into small finite elements, and for each of them we computed the electron-scattering optical depth, $\tau,$ along our sightline. We then computed the flux ${\cal F}$ for each rotational phase of the star:
\begin{equation}
  {\cal F} = \int_{0}^{\pi/2} I(\mu)\,\exp(-\tau(x,y))\,\mu\, d\theta
,\end{equation}
where $I(\mu)$ is the limb-darkened specific intensity, $\tau(x,y)$ the optical depth along the sightline starting on the coordinates $(x,y)$ on the projection of the stellar surface, and $\mu = \cos{\theta} = \sqrt{1 - \frac{x^2 + y^2}{R_*^2}}$. For the limb darkening, we tested both a linear relation, $I(\mu) = I_0\,[1 - u\,(1 - \mu)],$ and a so-called Power-2 relation, $I(\mu) = I_0\,[1 - g\,(1 - \mu^h)],$ with the $u$, $g,$ and $h$ parameters taken, respectively, from \citet{Cla17} and \citet{Cla22}.

\begin{table}
  \begin{center}
  \caption{Default CIR model parameters. \label{CIRparam}}
  \begin{tabular}{l c c}
    \hline
    Parameter & Value & Ref.\\
    \hline
    $v_{\rm rot}$ (km\,s$^{-1}$) & 100 \\
    $v_0$ (km\,s$^{-1}$) & 10 \\    
    $v_{\infty}$ (km\,s$^{-1}$) & 1900 & $[1]$\\
    $\frac{\dot{M}}{4\,\pi\,R_*\,v_{\infty}}$ (g\,cm$^{-2}$) & $2.21\,10^{-2}$ & $[1]$\\
    latitude range & $[-30^{\circ},30^{\circ}]$ & $[2]$\\
    $\rho_{\rm CIR}/\rho_{\rm wind}$ & 5.0 & $[2]$\\
    $u$ & 0.27 & $[3]$\\
    $g$ & 0.75 & $[4]$\\
    $h$ & 0.20 & $[4]$\\
    \hline
  \end{tabular}
  \tablefoot{References: $[1]$ \citet{Bou12}, $[2]$ \citet{Har00}, $[3]$ \citet{Cla17}, $[4]$ \citet{Cla22}.}
  \end{center}
\end{table}

\begin{figure*}
  \begin{minipage}{8.7cm}
    \resizebox{8.7cm}{!}{\includegraphics{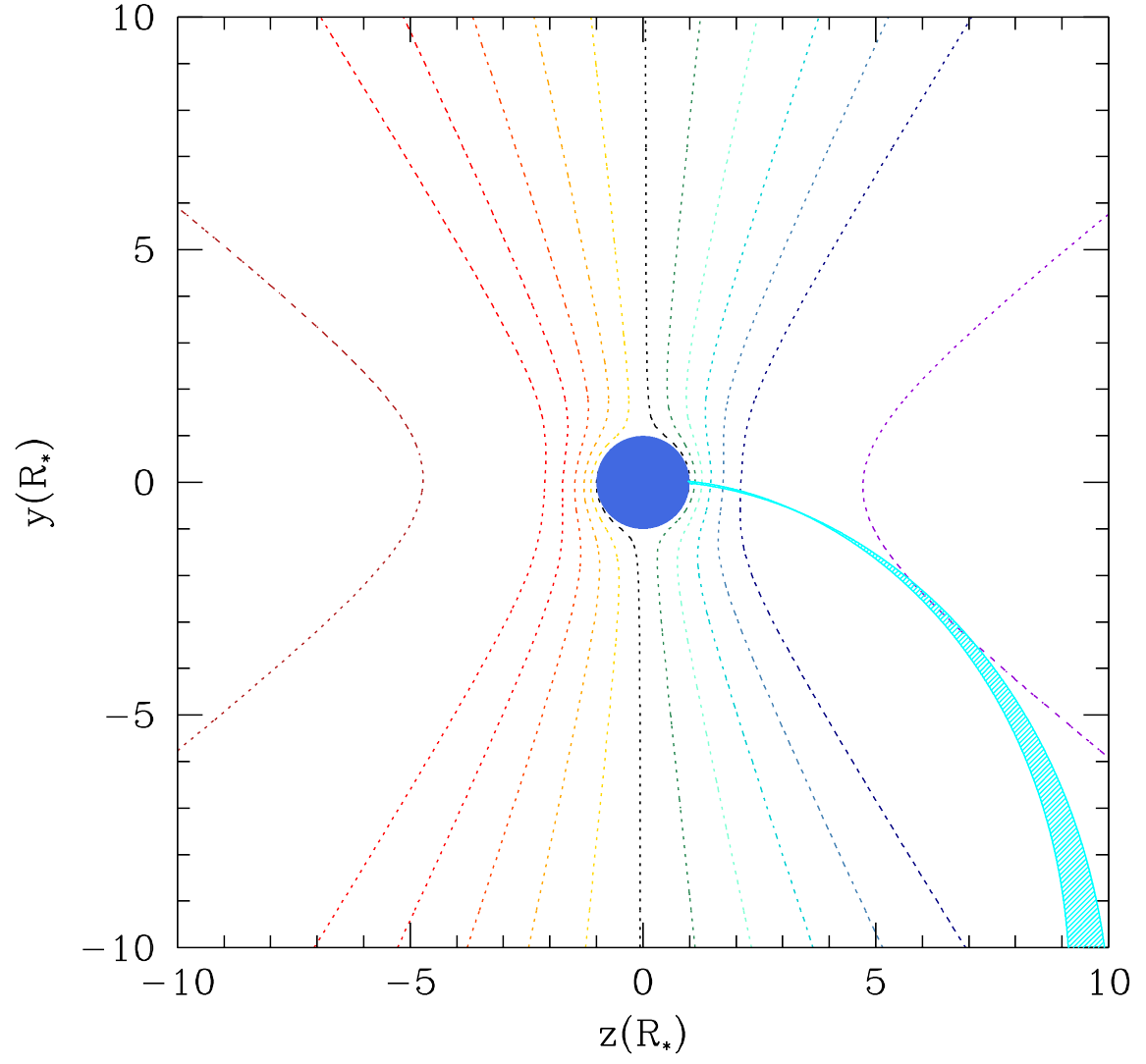}}
  \end{minipage}
  \hfill
  \begin{minipage}{8.7cm}
    \resizebox{8.7cm}{!}{\includegraphics{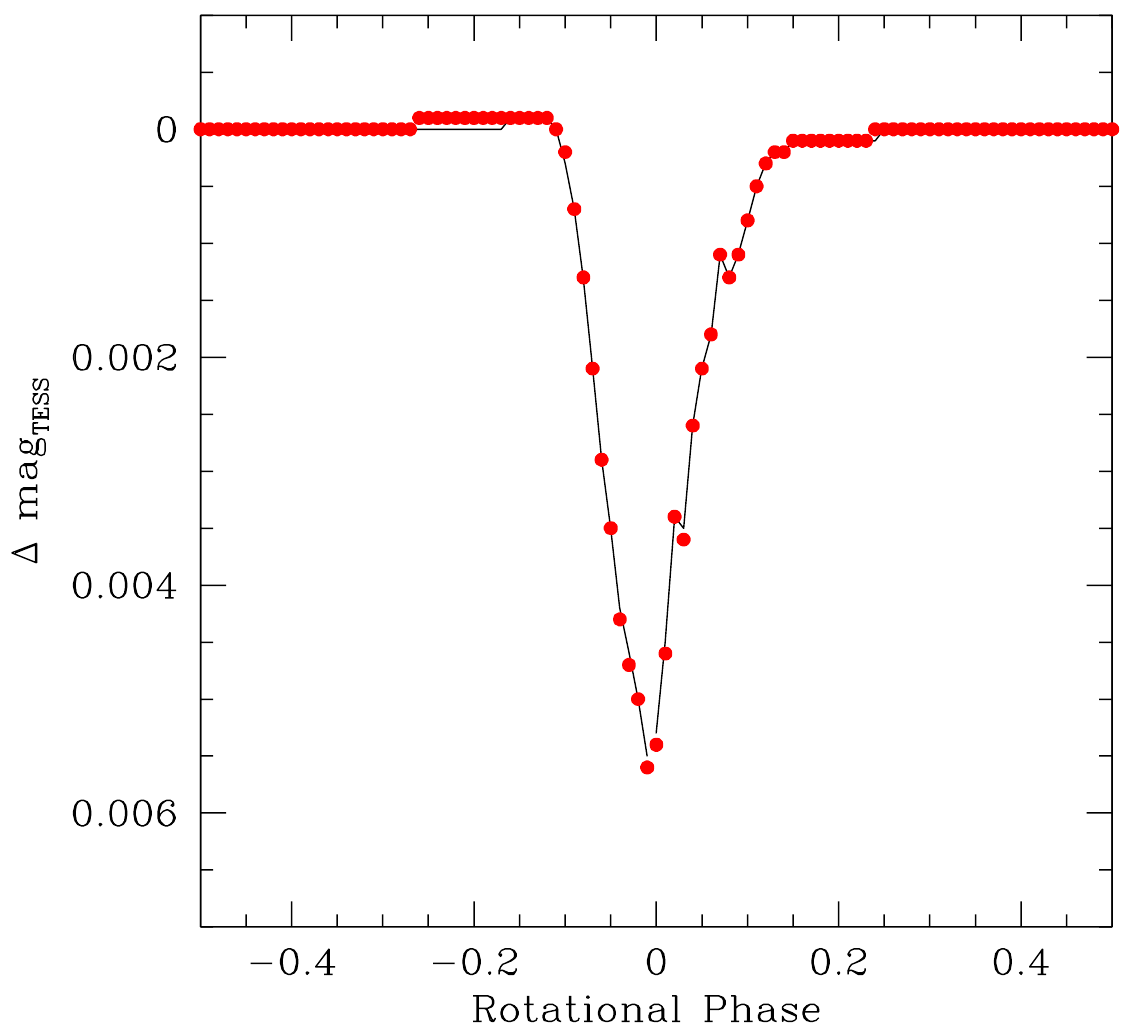}}
  \end{minipage}
  \caption{{\it Left}: Schematic representation of the CIR in the equatorial plane for our default model (see Table\,\ref{CIRparam}) at rotational phase 0.0. The observer is located at $z \rightarrow \infty$ and $y = 0$. The star rotates anti-clockwise. The dotted coloured contours represent iso-radial velocity surfaces with values of 1500, 1000, 800, 600, 400, 200, 0, -200, -400, -600, -800, -1000, and -1500\,km\,s$^{-1}$ (from left to right). {\it Right}: Predicted light curve due to electron scattering by the co-rotating CIR with our default model. The black curve yields the results for a Power-2 limb-darkening relation, whilst the red dots correspond to a linear limb darkening. \label{CIRplot}}
\end{figure*}

Figure\,\ref{CIRplot} shows a schematic representation of a CIR and the predicted light variations due to electron scattering. Regardless of the adopted limb-darkening relation, we find predicted light curves that are slightly asymmetric with a steeper descent towards the minimum followed by a more progressive recovery as the spiral rotates along with the star. This is unlike our observations, which rather indicate a more or less symmetrical trough (see Fig.\,\ref{Sect41}), although the shallow extension of the recovery phase could be hidden by the red noise variability, making it difficult to see the asymmetry in practice. However, more important differences exist. For our default model, the total duration of the event is about 0.2 times the rotational period. This is about three times less than the duration of the trough event in HD~192639, which lasts 3\,d over a 5\,d rotation period. Furthermore, another important difference concerns the depth of the event; our simulations with the default model yield a depth of only about 0.006\,mag, or about one dex smaller than observed.

We tested the influence of several model parameters. First, we varied the value of the wind column density, $\frac{\dot{M}}{4\,\pi\,R_*\,v_{\infty}}$, in the range of the various model atmosphere studies between $0.0185$\,g\,cm$^{-2}$ \citep{Haw21} and $0.0412$\,g\,cm$^{-2}$ \citep{Gor22}; we found that the maximum depth of the event varies roughly linearly with wind column density. With all other parameters unchanged, the depth reaches at most 0.01\,mag, which is still five times less than the observed value, for the largest column density. Neither the duration nor the shape of the event is sensitive to this parameter. As a second step, we investigated the impact of $v_{\infty}$. Keeping the rotational velocity at 100\,km\,s$^{-1}$, the wind velocity impacts the curvature of the spiral structure. A higher $v_{\infty}$ leads to a more radial CIR structure, increasing the maximum column density right above the spot responsible for the CIR. However, the shape of the light curve, which results from integration over the full stellar disc, remains essentially unchanged within the range of $v_{\infty}$ values (Table \ref{modelparam}).

One possibility to approach the observed depth of the trough is to increase the density contrast between the CIR and the ambient wind to a value of 20. In this case, with the wind-density parameter of \citet{Gor22}, our model predicts a depth of 0.045\,mag, similar to observations, but only with a duration of about 0.25 of the rotation cycle. Another possibility to increase the depth of the event is to assume that the spot covers a wider longitude range. Increasing the thickness of the CIR in the direction perpendicular to the spiral path by a factor of 10 leads to a similar depth to the observed trough and simultaneously increases the duration of the event to about 0.4 of the rotation cycle (closer but still shorter than observations). Finally, we doubled the latitude range covered by the spot to $[-60^{\circ},60^{\circ}]$. Compared to the default model, this only leads to a small ($\simeq 9$\%) increase of the depth. Therefore, the main parameters that rule the depth of the event are clearly the density contrast and the longitudinal extension of the CIR. Overall, it thus seems clear that quite extreme values for the properties of the CIR regions are required to reproduce the observations. Finally, we used the simple kinematic description of \citet{Cra96} and \citet{Ful97} to express the RV of the wind in the equatorial plane as
\begin{equation}
  RV = -v_{\infty}\,\left[w_0 + (1-w_0)\,\left(1-\frac{R_*}{r}\right)\right]\,\cos{\phi} + \frac{v_{\rm rot}\,R_*}{r}\,\sin{\phi}
  \label{CIRvr}
,\end{equation}
where $\phi$=0.0 corresponds to the photometric minimum and the footpoint of the CIR being aligned with the observer's sightline.

Considering that the observations of the feature in the He\,{\sc i} and H\,{\sc i} are due to material from a single distance $r_1$ in the CIR, we then computed the RV of the absorption as a function of rotation phase. Assuming a rotational period of 5\,d, the TIGRE observations taken near the photometric minimum would cover rotational phases between 0.04 and 0.10. The observed RVs of the blueshifted absorption (between $\sim -200$ and $-350$\,km\,s$^{-1}$) are consistent with a formation region in the CIR near 1.1 -- 1.2\,R$_*$. However, there is one important caveat. Assuming the absorption to arise at a specific distance from the star and assuming a continuous ejection of material into the CIR, Eq.\,\ref{CIRvr} predicts the ensuing absorption to move redwards with time after the light-curve minimum instead of bluewards as observed (see Fig.\,\ref{DAC_RV}). This is due to the distortion of the iso-RV surfaces (see the left panel of Fig.\,\ref{CIRplot}), which is especially pronounced at low RV and near the stellar surface. This discrepancy suggests that Eq.\,\ref{CIRvr} is too simplistic or that some of the assumptions made above are inappropriate. Together with the above conclusion on the extreme values required to reproduce the event, the velocity discrepancy argues against a 'normal' CIR as the source of the dimming event.

The model that we used here is relatively simple, both in terms of radiative transfer (electron scattering as the sole source of opacity) and wind dynamics (simplified velocity law). Given the wide bandwidth of the {\it TESS} cameras, the photometry is largely dominated by the continuum. Including line opacity is therefore unlikely to significantly change our conclusions about the synthetic light curve. As long as electron scattering remains the main source of opacity, clumping will not change our results either. Indeed, electron scattering has a linear dependence on density and remains thus insensitive to clumping. A more sophisticated treatment of radiative transfer is also unlikely to make a significant difference. In fact, despite the increase of the column density during the trough event, the wind remains optically thin. Our assumptions on the dynamics of the CIR (compression ratio, velocity law, etc.) likely have a stronger impact on the properties of the synthetic light curve. A more complex velocity field along with a detailed ionisation structure of the wind and the CIR could lead to a better agreement with the observed bluewards migration of the absorption component. Elaborating such a model is beyond the scope of the present study.  

\section{Conclusion}
Thanks to a high-cadence photometric and spectroscopic dataset, the variability of HD~192639 is now clarified. As this star is often considered a prototypical supergiant, our study has a general interest for the knowledge of O-type supergiants. The observed variability can be split in three categories. First, low-frequency stochastic changes (also known as red noise) are detected both in photometry and spectroscopy. This detection supports the usual consideration of its ubiquity in massive hot stars, including evolved ones.
 
Second, a $\sim 5$\,d timescale is repeatedly detected, both in photometry and spectroscopy, even in datasets taken years apart. This timescale most probably corresponds to the rotation period of the star. Overall, many of the properties of this recurrent variability are consistent with a CIR driven by a spotted photosphere.

Finally, a dimming that is both longer (3\,d) and stronger (0.05\,mag) than any other one is also detected in one of the seven TESS light curves. Simultaneous spectroscopy reveals enhanced emission and/or absorptions in He\,{\sc i} and Balmer lines that slowly shift with time. This suggests an episodic ejection, whose material then slowly blends with the circumstellar environment. To the best of our knowledge, such an event has not been reported before for an O-star.

Several questions then arise.
The dimming amplitude corresponds to a doubling of the wind column density, which is enormous. An important question is what triggers the ejection of such a large amount of mass. The beating of several non-radial pulsations seems unlikely, as no such pulsations were seen in our spectroscopic and photometric analyses of HD~192639. Alternatively, one could consider the scenario where the star features spots due to localised magnetic fields, which is used to explain the rotational changes. In such a configuration, magnetic reconnections could lead to a scaled-up version of solar coronal mass ejection events. However, in this case, one would probably expect the ejected material to be quite hot, unlike what we observe here. Indeed, the associated spectroscopic signature is only observable in He\,{\sc i} and H\,{\sc i} lines, not He\,{\sc ii} lines. 

Another question concerns the frequency of such events. Enhanced absorptions as in CIRs and DACs have recurrence times of the order of a few days to about two weeks \cite[e.g.][]{Kap99,Lob08}. Whilst the {\it TESS} light curves of the seven sectors unveil about five other dips, none of them was as extreme in terms of duration and depth as the trough itself. Therefore, the event recorded for HD~192639 clearly occurs less frequently. Furthermore, it involves a significantly larger quantity of material than conventional DACs. An important point is also that it does not seem to last for long: whilst the {\it TESS} light curve of Sector 41 shows another dip about five days after the trough, it is much shallower and did not produce a signature in the spectral lines. Therefore, the mechanism that caused the ejection must have weakened or even nearly vanished within one rotational period.

The last question concerns the impact of such events on the stellar feedback and evolution. Indeed, such events may trace an overlooked aspect of the mass-loss phenomenon. However, a single detection in a single star is of course not enough to assess the importance of such ejections. While it is impossible to organise simultaneous high-cadence photometric and spectroscopic campaign for each O-type star in the sky, a first step forward could be to investigate the existing {\it TESS} data of presumably single O-stars (to avoid contamination by binary interaction effects) and search for trough-like events.

\begin{acknowledgements}
YN acknowledges support from the Fonds National de la Recherche Scientiﬁque (Belgium). This work used data collected by the {\it TESS} mission, which are publicly available from the Mikulski Archive for Space Telescopes (MAST). Funding for the TESS mission is provided by NASA’s Science Mission directorate. ADS and CDS were used during this research. We thank the anonymous referee for remarks that helped us improve our manuscript.
\end{acknowledgements}

\bibliographystyle{aa}
  \bibliography{mybiblio}
  \begin{appendix}
  \section{Time frequency diagrams of other {\it TESS} sectors \label{AppFourier}}
\begin{figure}[h]
  \begin{center}
    \resizebox{7.3cm}{!}{\includegraphics{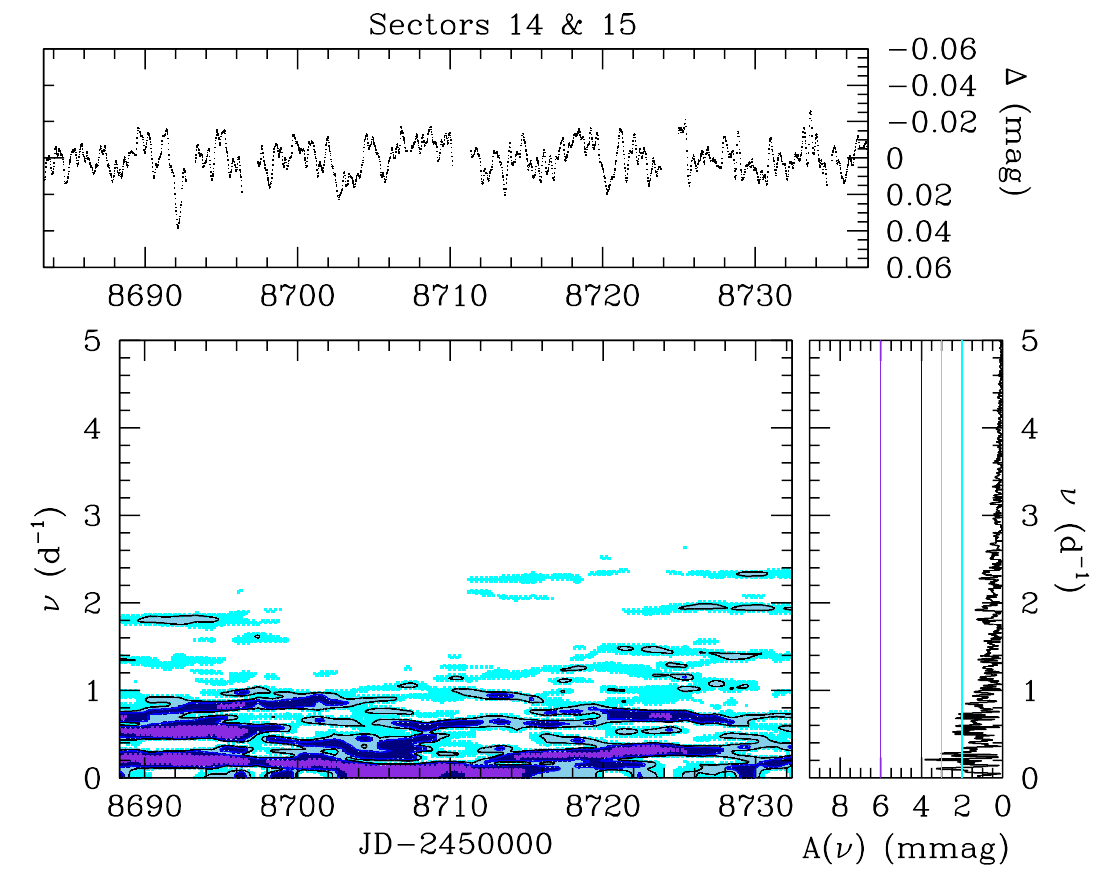}}
  \end{center}  
  \caption{Same as Fig.\,\ref{Sect41} but for Sectors 14 \& 15. \label{Sect1415}}
\end{figure}

\begin{figure}[h]
  \begin{center}
    \resizebox{7.3cm}{!}{\includegraphics{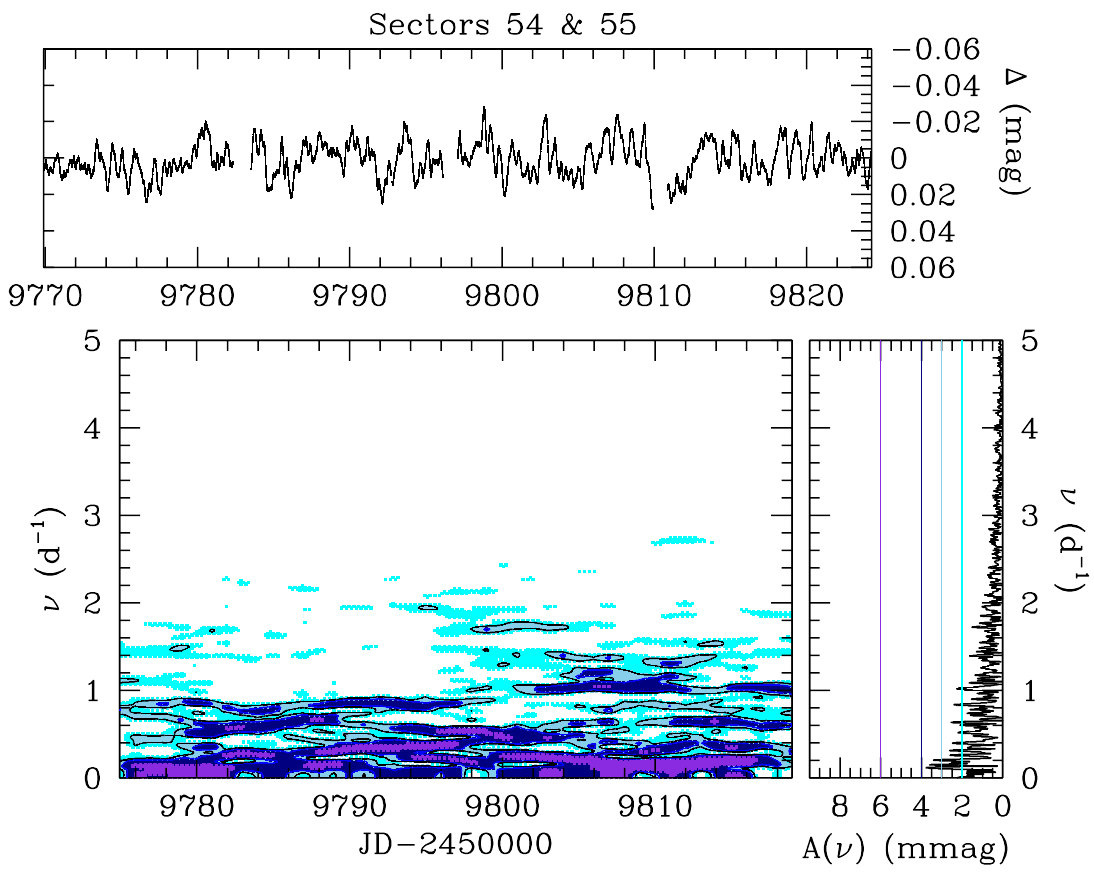}}
  \end{center}
  \caption{Same as Fig.\,\ref{Sect41} but for Sectors 54 \& 55.\label{spevol3}}
  \end{figure}

\begin{figure}[h]
  \begin{center}
    \resizebox{7.3cm}{!}{\includegraphics{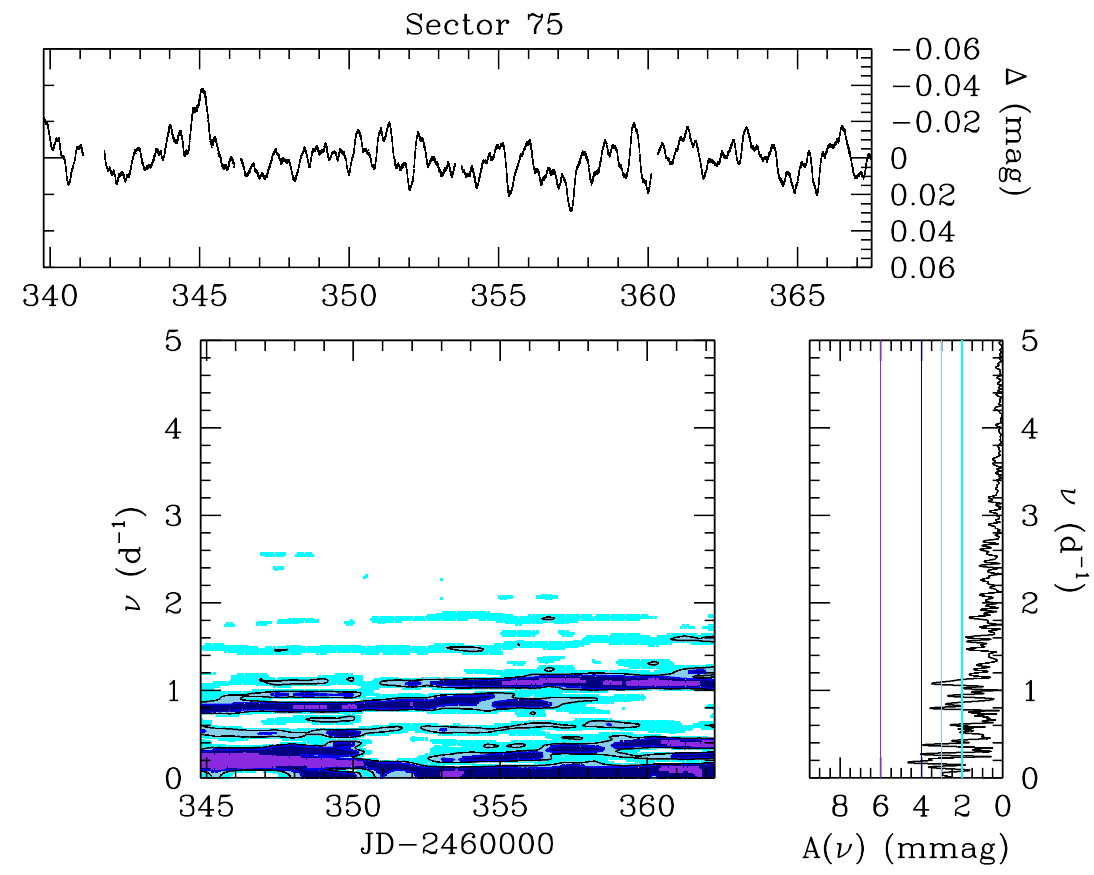}}
  \end{center}
  \caption{Same as Fig.\,\ref{Sect41} but for Sector 75.\label{spevol4}}
\end{figure}

\begin{figure}[h]
  \begin{center}
    \resizebox{7.3cm}{!}{\includegraphics{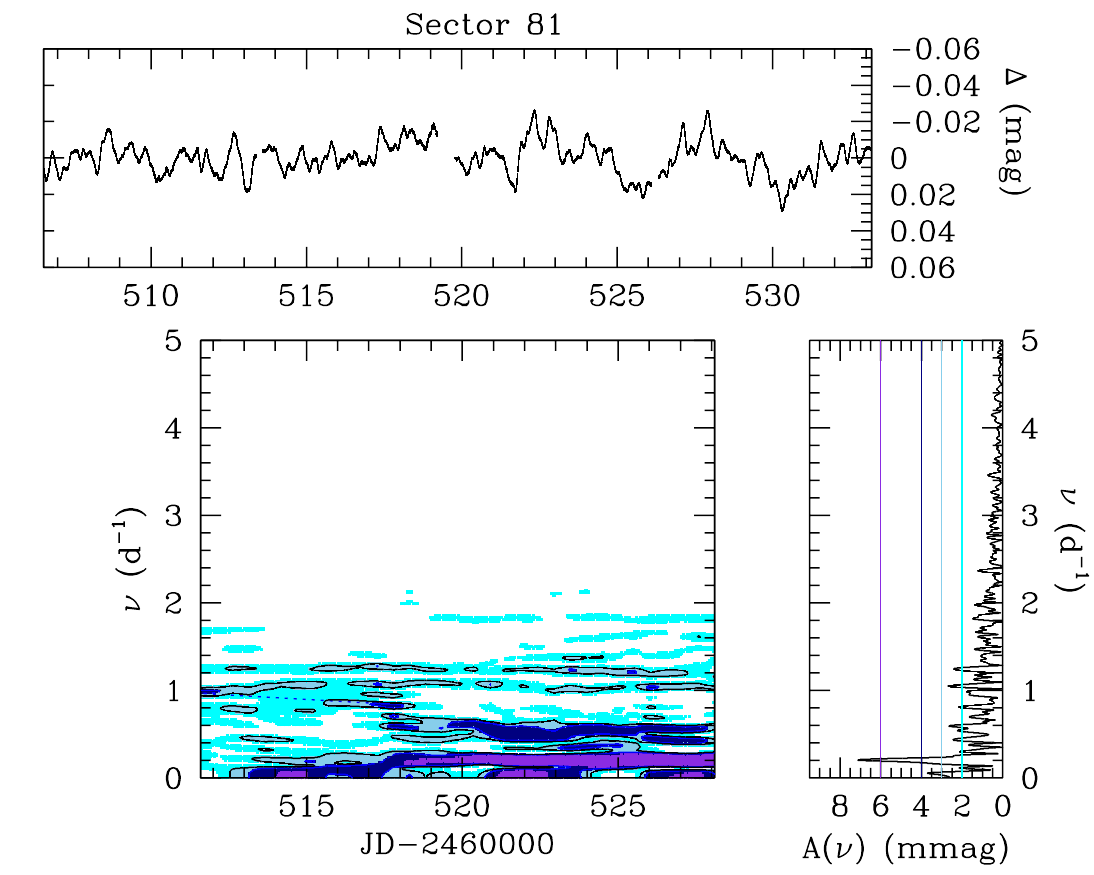}}
  \end{center}
  \caption{Same as Fig.\,\ref{Sect41} but for Sector 81.\label{spevol5}}
\end{figure}

  Figures\,\ref{Sect1415}-\ref{spevol5} display the time frequency diagrams of the {\it TESS} Sectors 14 \& 15, 54 \& 55, 75 and 81. The notations are the same as for Fig.\,\ref{Sect41}.

\section{Projected rotational velocity determination \label{AppRotation}}
\begin{figure}[h!]
  \begin{center}
    \begin{minipage}{4.4cm}
      \resizebox{4.5cm}{!}{\includegraphics{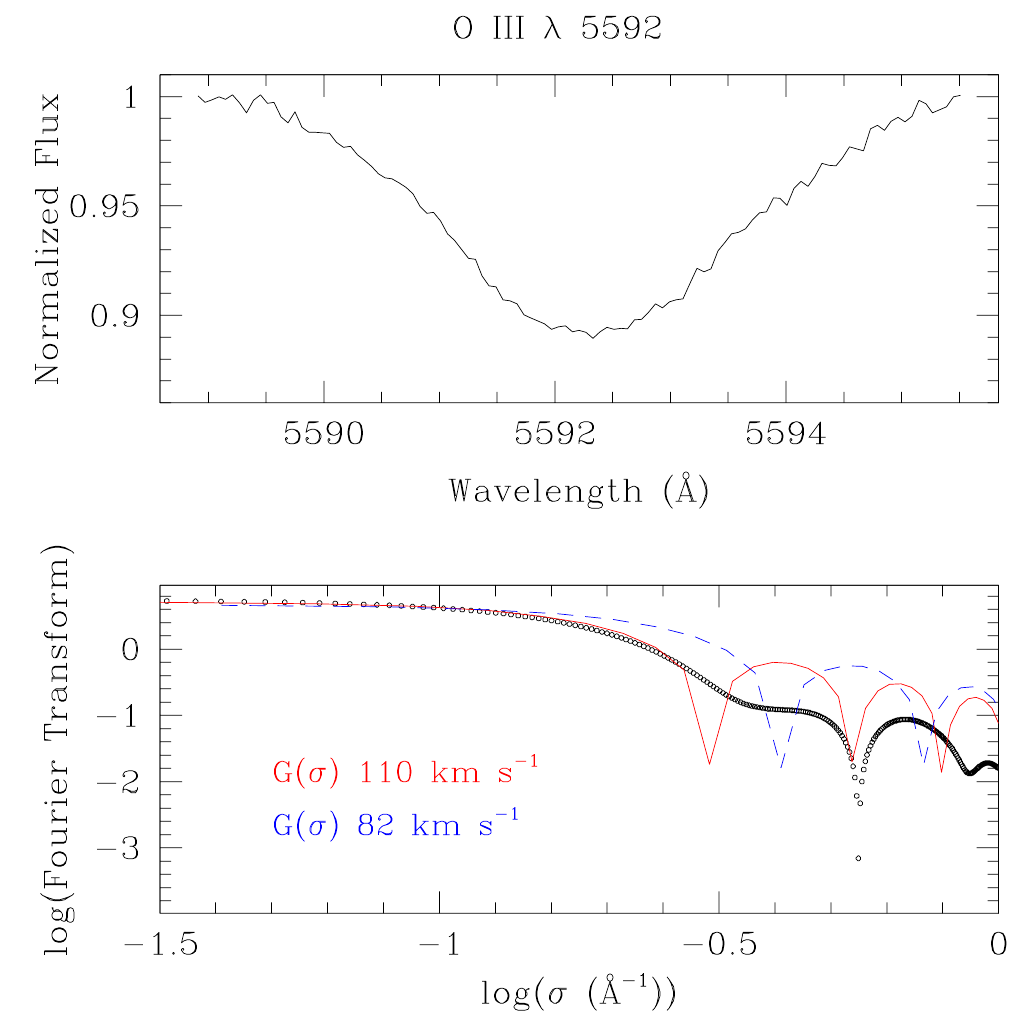}}
    \end{minipage}
    \hfill
    \begin{minipage}{4.4cm}
      \resizebox{4.5cm}{!}{\includegraphics{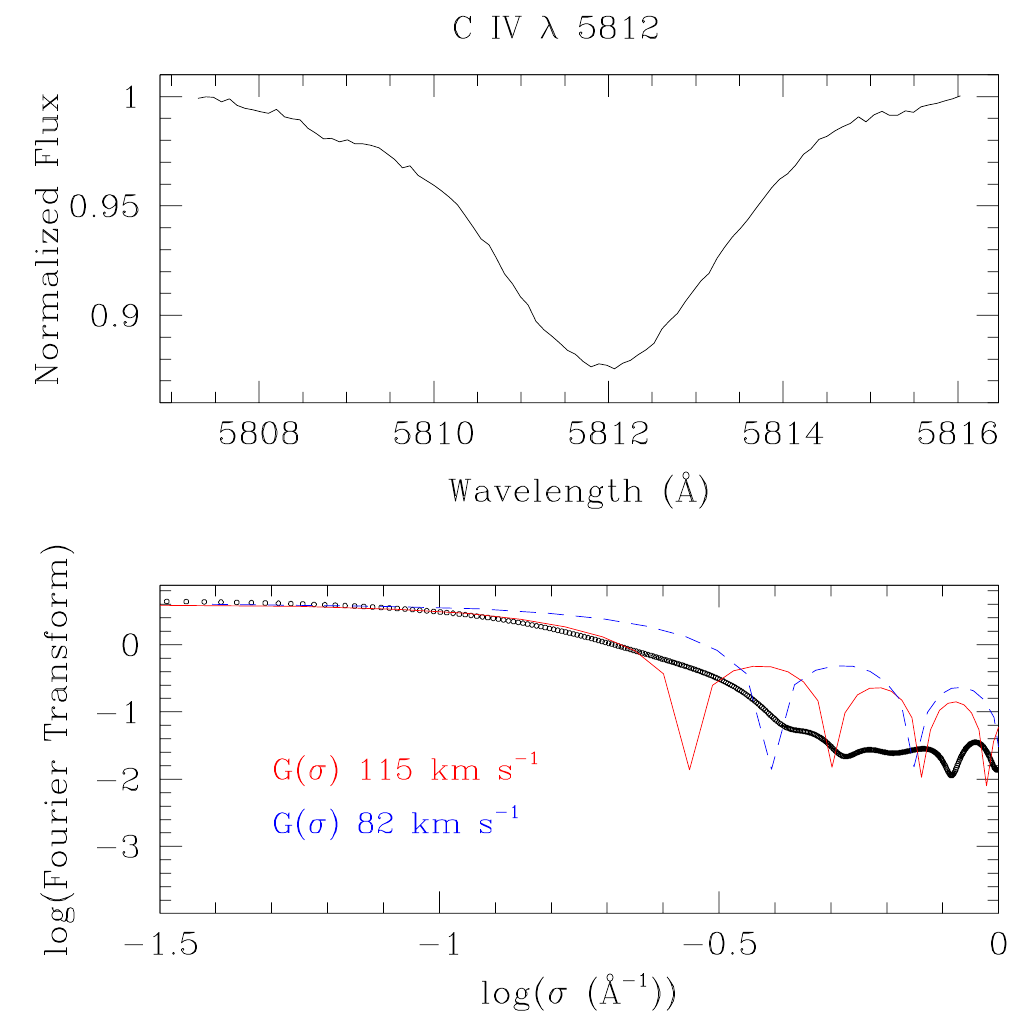}}
    \end{minipage}
  \end{center}
 \caption{Determination of $v\,\sin{i}$ using the Fourier transform method on the O\,{\sc iii} $\lambda$\,5592 (left panel) and C\,{\sc iv} $\lambda$\,5812 (right panel) lines. For each line, the top subpanel illustrates the mean line profile, whilst the bottom subpanel provides the Fourier transform of the observed profile (open circles) and the Fourier transform of a rotational broadening function for our best-fitting value (red curve) and for the value quoted by \citet[][blue curve]{Hol22}. \label{vsini}}
\end{figure}

Figure\,\ref{vsini} illustrates the application of the Fourier method to the O\,{\sc iii} $\lambda$\,5592 and C\,{\sc iv} $\lambda$\,5812 lines. Both lines are located in a part of the spectral domain where the sensitivity of the TIGRE/HEROS configuration is low. We thus combined all 108 spectra for this analysis. One should keep in mind that \citet{Ful96} detected lpv in the C\,{\sc iv} $\lambda$\,5812 line. 

\section{Temporal variance spectra of other spectral lines \label{AppTVS}}
\begin{figure*}[h!]
 \begin{minipage}{6cm}
  \begin{center}
    \resizebox{6cm}{!}{\includegraphics{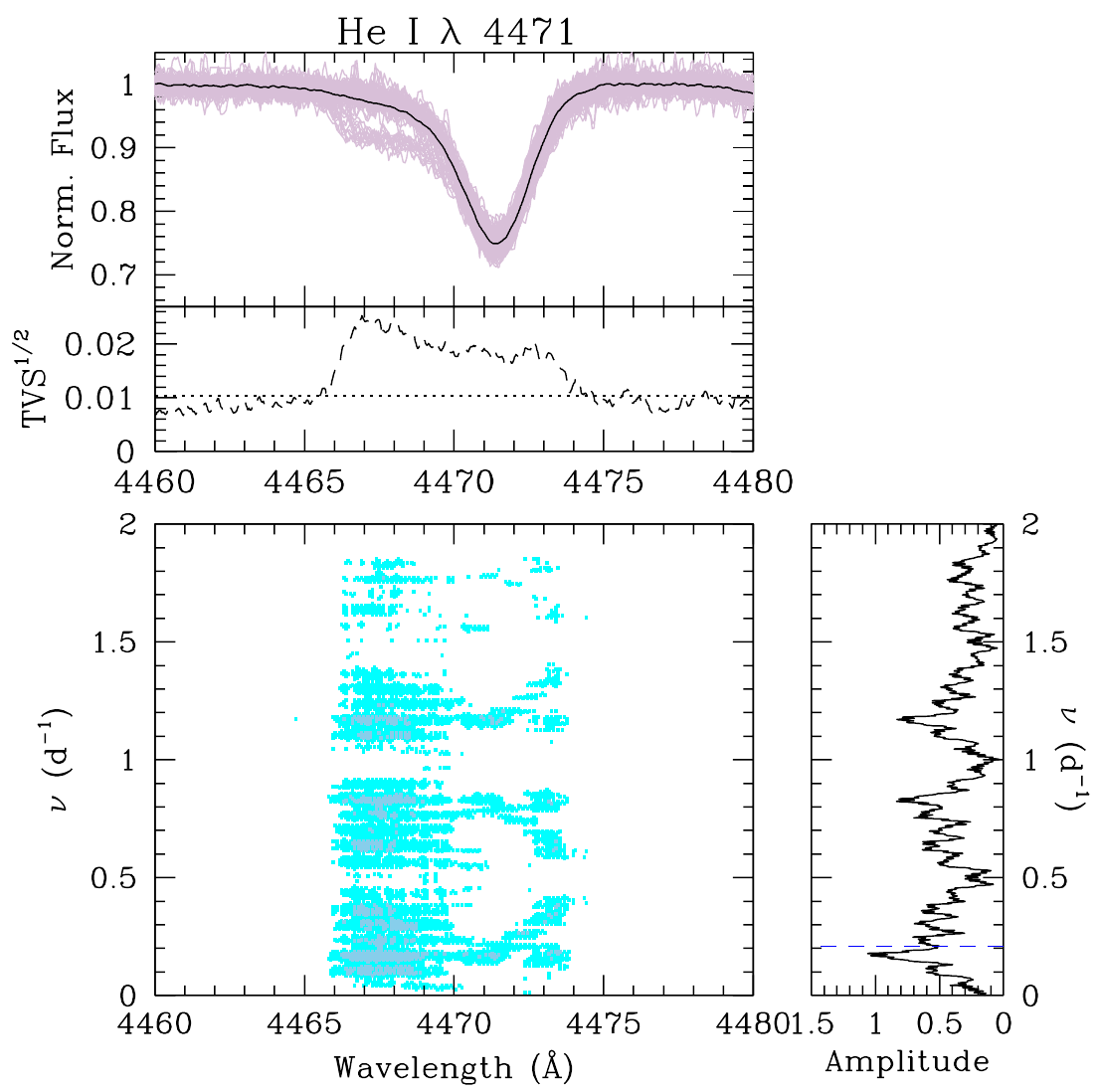}}
  \end{center}
 \end{minipage}
 \hfill
 \begin{minipage}{6cm}
  \begin{center}
    \resizebox{6cm}{!}{\includegraphics{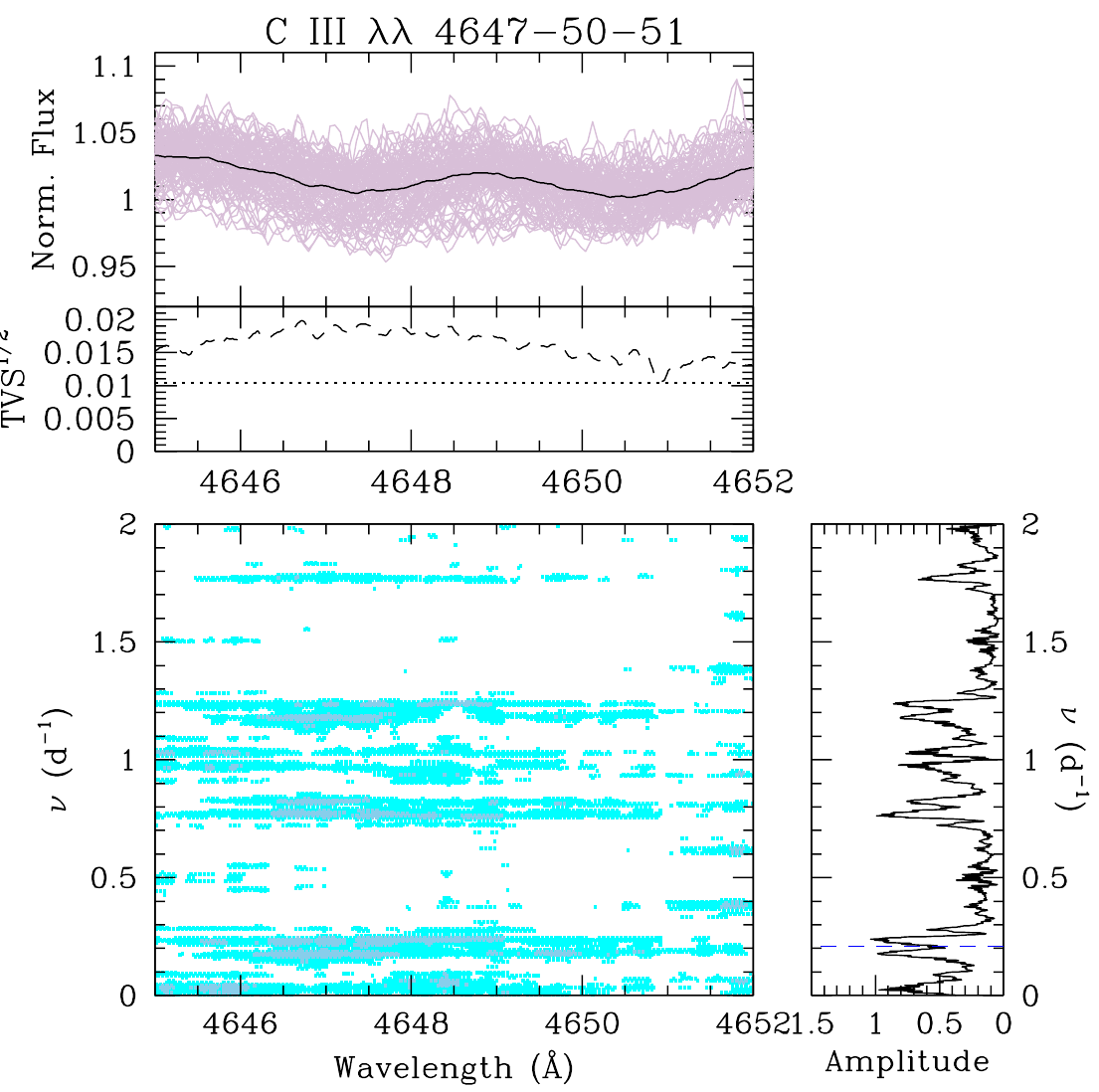}}
  \end{center}
 \end{minipage}
 \hfill
 \begin{minipage}{6cm}
  \begin{center}
    \resizebox{6cm}{!}{\includegraphics{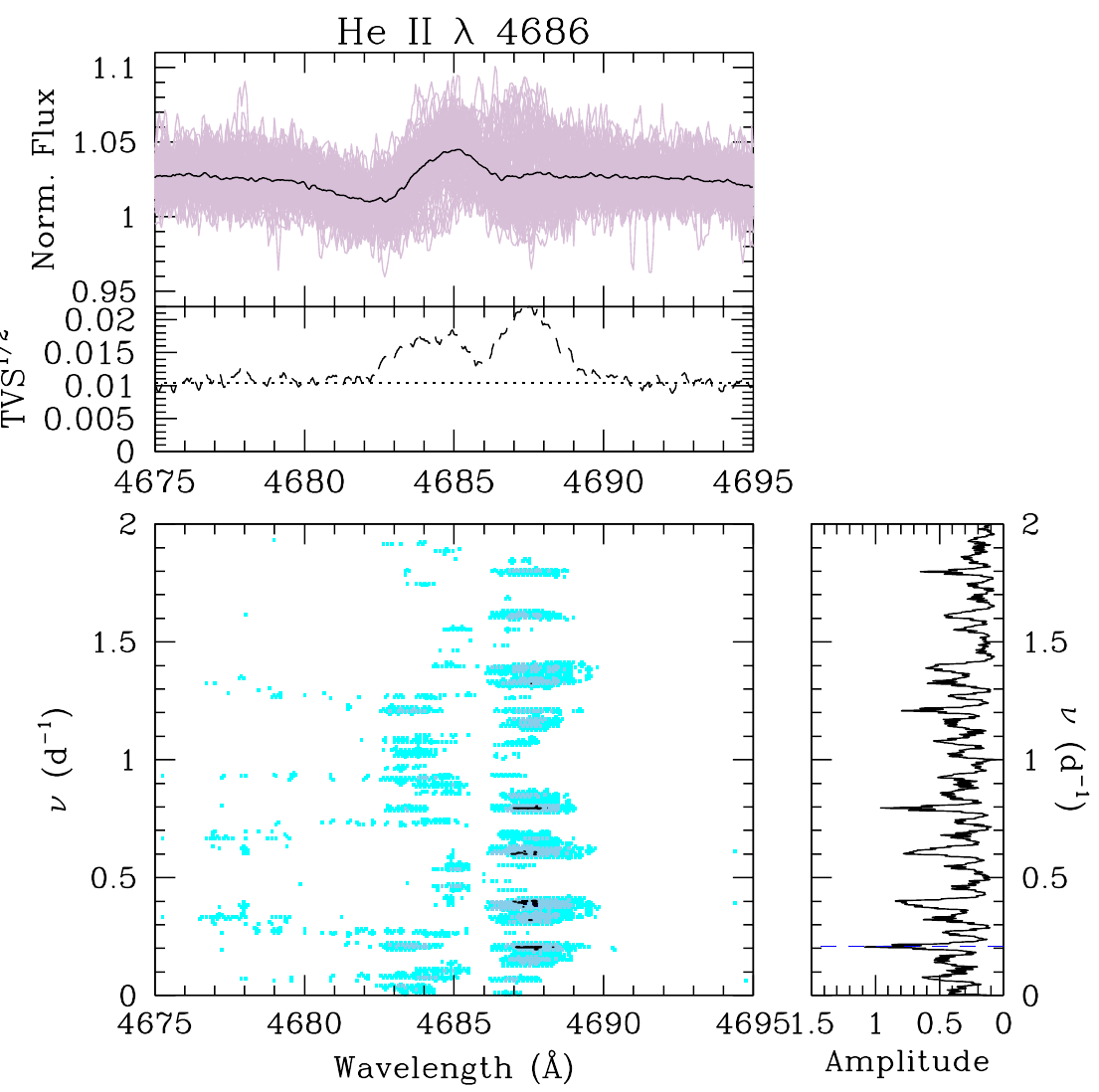}}
  \end{center}
 \end{minipage}
 \begin{minipage}{6cm}
  \begin{center}
    \resizebox{6cm}{!}{\includegraphics{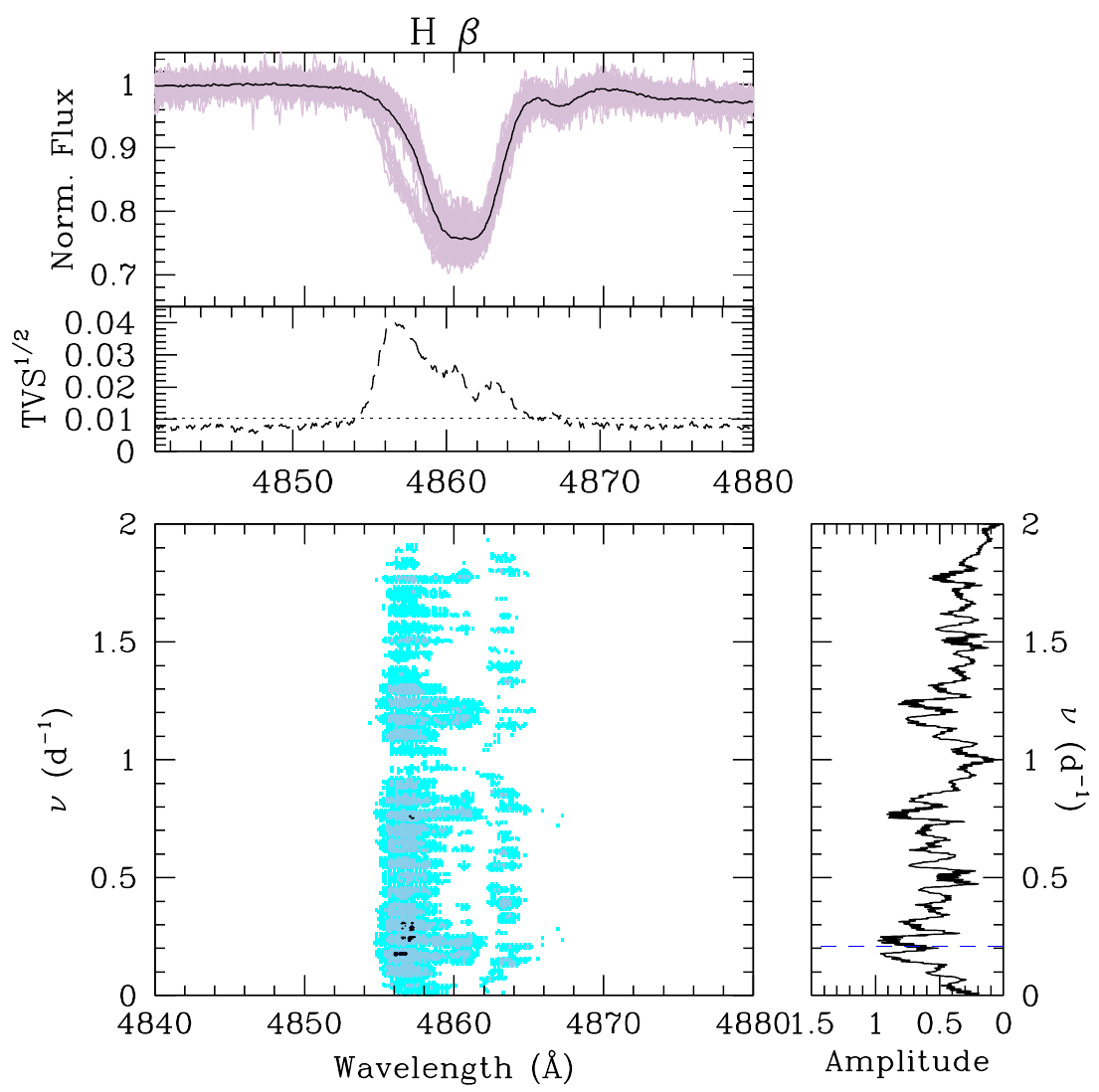}}
  \end{center}
 \end{minipage}
 \hfill
 \begin{minipage}{6cm}
  \begin{center}
    \resizebox{6cm}{!}{\includegraphics{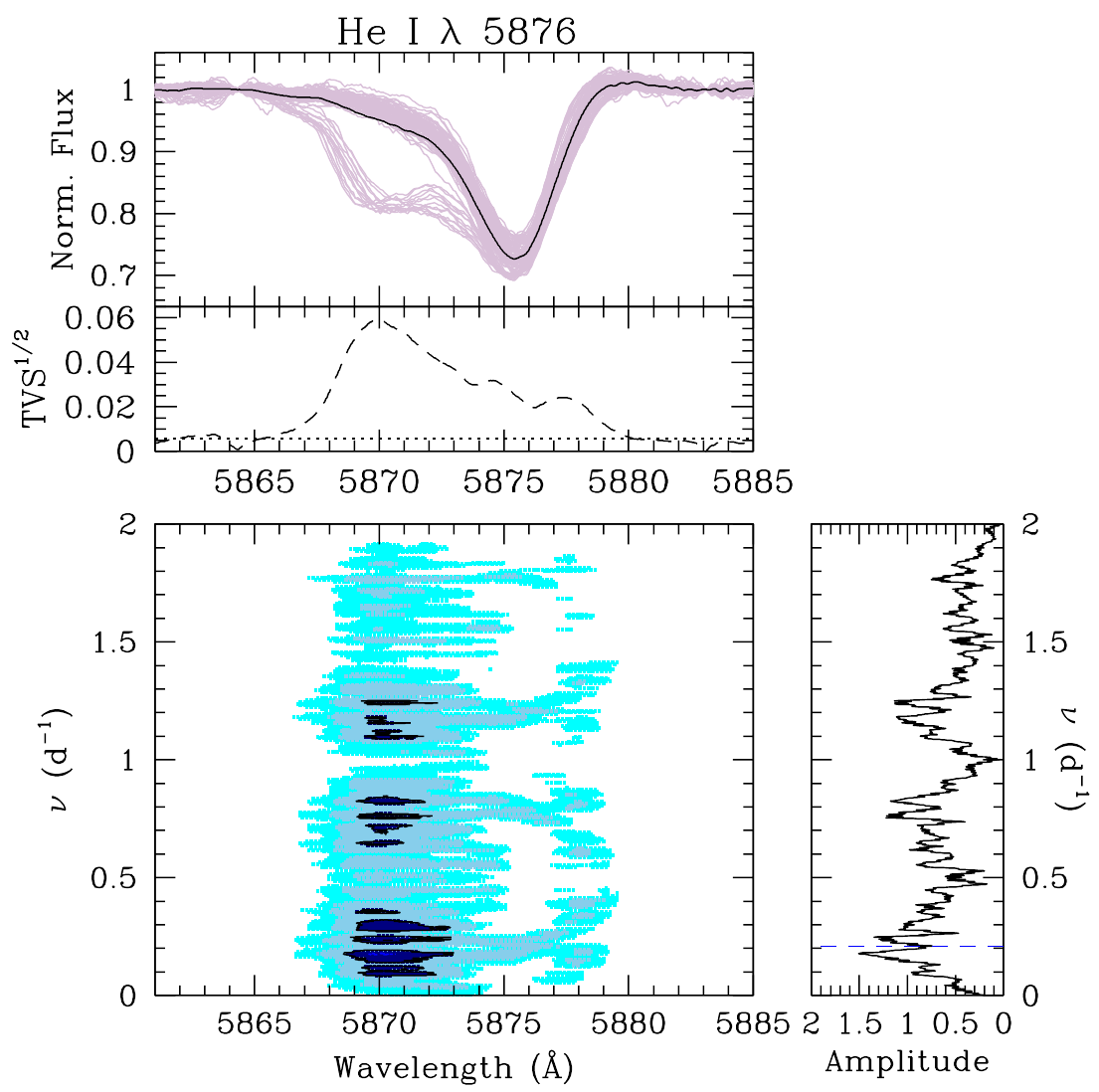}}
  \end{center}
 \end{minipage}
 \hfill
 \begin{minipage}{6cm}
  \begin{center}
    \resizebox{6cm}{!}{\includegraphics{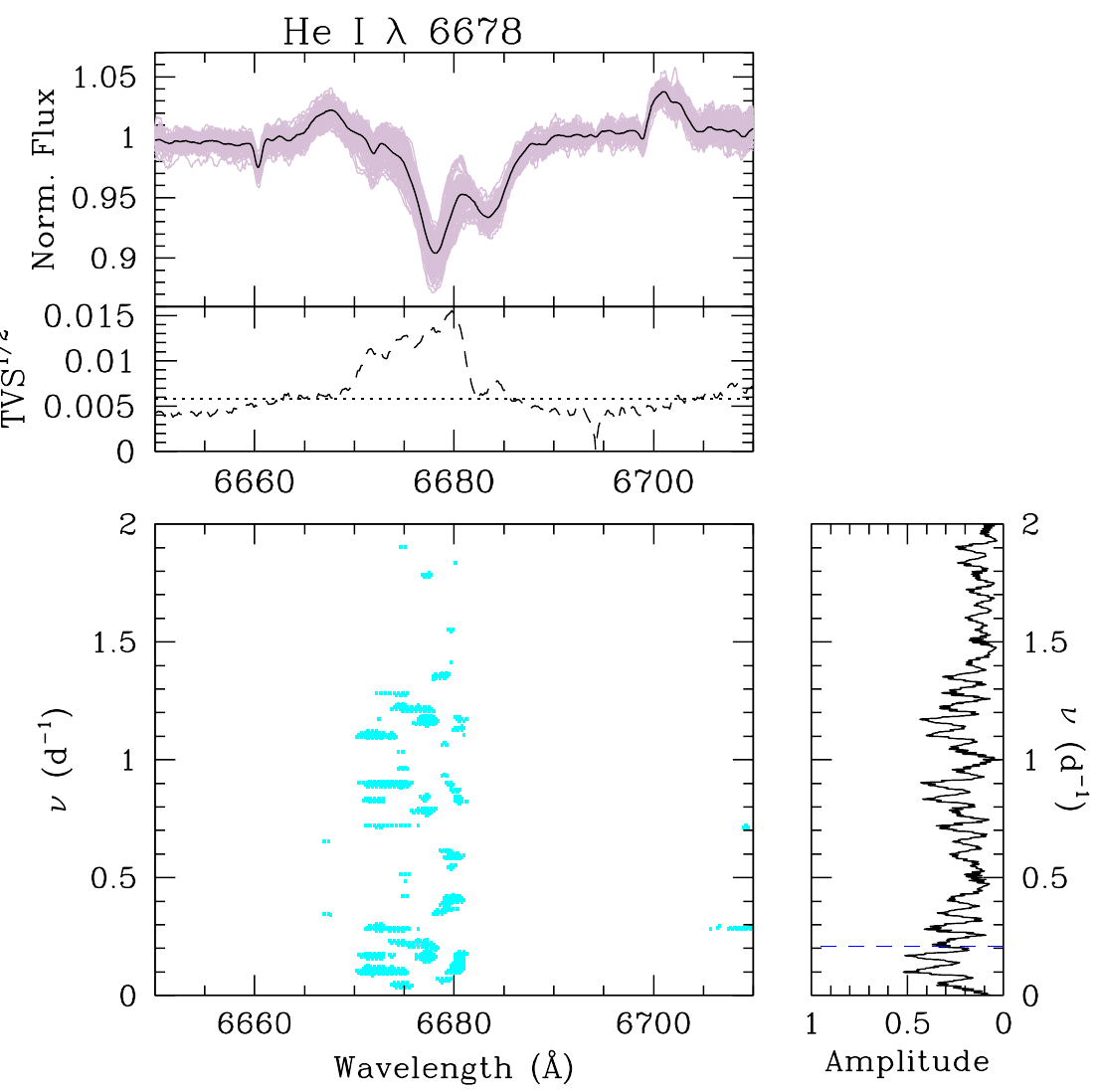}}
  \end{center}
 \end{minipage}
 \caption{Same as Fig.\,\ref{TVS_Ha} for some selected spectral lines in our TIGRE/HEROS data of HD~192639. The colours in the bottom left panel indicate ratios between the amplitude of the periodogram and the 1\% significance level of 0.5 (cyan), 1.0 (skyblue), 2.0 (navyblue) and 3.0 (blueviolet).\label{TVS_bis}}
\end{figure*}

Figure\,\ref{TVS_bis} provides the TVS and Fourier periodograms of variable lines in the TIGRE/HEROS spectra of HD~192639. The notations are the same as for Fig.\,\ref{TVS_Ha}.

\FloatBarrier

\section{Journal of spectroscopic observations \label{AppJournal}}
Table\,\ref{Journal} provides the journal of our 108 TIGRE/HEROS observations of HD~192639 along with the RVs that were measured for individual absorption lines.
\begin{table*}[h!]
  \caption{Journal of the TIGRE spectroscopic observations of HD~192639 \label{Journal}}
  \begin{center}
  \begin{tabular}{c c r r r r r r}
\hline
HJD-2459000 & S/N & \multicolumn{6}{c}{RV (km\,s$^{-1}$)} \\
& & He\,{\sc ii} $\lambda$\,4200 & H$\gamma$ & He\,{\sc i} $\lambda$\,4471 & He\,{\sc ii} $\lambda$\,4542 & H$\beta$ & He\,{\sc ii} $\lambda$\,5412 \\
\hline
309.9826 & 109 & $-17.42$  & $-38.26$  & $-21.32$  & $-10.63$  & $-53.16$  & $ -9.31$ \\
313.9514 & 241 & $ -1.00$  & $-32.39$  & $ -7.38$  & $  1.52$  & $-40.95$  & $  1.27$ \\
315.9551 & 178 & $ -7.78$  & $-39.44$  & $-18.44$  & $ -6.54$  & $-62.04$  & $ -5.48$ \\
317.9429 & 187 & $  4.28$  & $-23.83$  & $ -6.24$  & $  2.11$  & $-35.64$  & $  3.10$ \\
319.9389 & 157 & $ -1.71$  & $-43.58$  & $-20.52$  & $ -7.86$  & $-67.28$  & $  2.11$ \\
321.9320 & 114 & $ -5.57$  & $-30.25$  & $ -4.43$  & $ -4.29$  & $-43.48$  & $  9.31$ \\
323.9191 & 141 & $  2.28$  & $-27.01$  & $-10.53$  & $ -0.66$  & $-26.02$  & $  2.94$ \\
325.9180 & 188 & $ -0.79$  & $-38.13$  & $-11.46$  & $  3.96$  & $-55.87$  & $  2.22$ \\
328.9052 & 179 & $ -8.07$  & $-38.33$  & $ -8.51$  & $ -4.82$  & $-54.21$  & $  1.16$ \\
334.9340 & 166 & $ -8.57$  & $-44.89$  & $-20.92$  & $ -3.04$  & $-66.73$  & $  9.20$ \\
336.9532 & 100 & $ -1.78$  & $-29.22$  & $ -3.02$  & $  1.19$  & $-47.18$  & $  2.77$ \\
416.7199 & 123 & $ -5.14$  & $-33.84$  & $-18.37$  & $  3.63$  & $-50.45$  & $  0.44$ \\
416.7408 & 128 & $ 11.21$  & $-42.89$  & $-11.53$  & $  1.06$  & $-57.35$  & $  7.98$ \\
416.7618 & 158 & $ -6.14$  & $-32.53$  & $-10.26$  & $  0.40$  & $-52.17$  & $ 14.02$ \\
416.7827 & 151 & $ -5.21$  & $-36.26$  & $-14.62$  & $ -0.92$  & $-55.32$  & $  9.69$ \\
418.7537 & 208 & $ -2.57$  & $-24.17$  & $ -5.70$  & $ -1.25$  & $-42.61$  & $  7.15$ \\
418.7746 & 308 & $ -2.43$  & $-27.08$  & $ -7.91$  & $ -1.19$  & $-41.56$  & $  9.86$ \\
418.7956 & 241 & $ -2.86$  & $-27.49$  & $ -8.31$  & $ -1.78$  & $-44.89$  & $  2.94$ \\
419.6379 & 158 & $ -3.28$  & $-27.21$  & $ -6.97$  & $  2.31$  & $-37.86$  & $  6.54$ \\
420.6673 & 179 & $ -5.57$  & $-27.01$  & $-10.59$  & $  1.12$  & $-41.81$  & $  1.33$ \\
420.6883 & 263 & $ -3.78$  & $-26.94$  & $ -6.44$  & $ -4.36$  & $-41.50$  & $  4.65$ \\
420.7092 & 273 & $-11.92$  & $-29.98$  & $-11.13$  & $ -1.58$  & $-41.50$  & $  2.33$ \\
420.7301 & 382 & $ -5.14$  & $-30.53$  & $ -8.98$  & $ -3.76$  & $-42.24$  & $ -0.44$ \\
420.7511 & 250 & $ -7.14$  & $-30.11$  & $ -9.32$  & $ -2.57$  & $-44.40$  & $ -3.49$ \\
420.7720 & 239 & $-10.56$  & $-27.42$  & $-11.06$  & $ -2.90$  & $-43.48$  & $ -1.00$ \\
420.7930 & 241 & $ -7.85$  & $-29.22$  & $-12.07$  & $ -5.02$  & $-46.19$  & $ -2.11$ \\
420.8139 & 276 & $ -7.35$  & $-29.56$  & $-12.20$  & $ -5.28$  & $-43.91$  & $ -3.77$ \\
421.6304 & 189 & $ -8.85$  & $-42.89$  & $-18.10$  & $ -8.12$  & $-60.81$  & $-10.30$ \\
421.6513 & 233 & $-16.42$  & $-42.62$  & $-20.38$  & $-11.55$  & $-60.31$  & $ -8.97$ \\
421.6722 & 195 & $-17.27$  & $-45.31$  & $-24.27$  & $ -5.61$  & $-64.69$  & $-13.02$ \\
421.7341 & 173 & $-16.70$  & $-47.66$  & $-24.00$  & $ -9.44$  & $-68.02$  & $-10.80$ \\
421.7550 & 243 & $-20.13$  & $-42.82$  & $-21.86$  & $-10.83$  & $-69.99$  & $ -9.31$ \\
421.7759 & 182 & $-14.63$  & $-43.31$  & $-18.97$  & $ -9.37$  & $-68.02$  & $-12.69$ \\
421.7969 & 138 & $ -8.92$  & $-46.48$  & $-20.25$  & $ -8.05$  & $-67.34$  & $ -3.55$ \\
421.8178 & 185 & $-10.92$  & $-40.96$  & $-20.72$  & $ -6.01$  & $-64.51$  & $ -3.99$ \\
422.6044 & 161 & $  5.42$  & $-31.08$  & $ -8.25$  & $  7.20$  & $-57.41$  & $  7.87$ \\
422.6254 & 142 & $  2.71$  & $-32.74$  & $ -7.24$  & $  7.53$  & $-58.15$  & $ 11.86$ \\
422.6797 & 106 & $ -1.14$  & $-34.53$  & $ -2.82$  & $  3.37$  & $-57.23$  & $  6.48$ \\
422.7007 & 116 & $  3.85$  & $-29.91$  & $ -3.69$  & $  2.77$  & $-55.87$  & $  5.76$ \\
422.7216 & 153 & $ -0.36$  & $-33.91$  & $ -6.97$  & $  3.17$  & $-55.93$  & $  4.93$ \\
422.7426 & 144 & $ -0.36$  & $-33.91$  & $ -7.51$  & $ -1.06$  & $-55.81$  & $  8.03$ \\
422.7635 & 166 & $ -0.14$  & $-34.74$  & $ -7.71$  & $  0.66$  & $-59.88$  & $  7.04$ \\
422.7844 & 126 & $ -1.78$  & $-33.15$  & $ -8.65$  & $ -0.46$  & $-61.05$  & $  5.21$ \\
422.8054 & 115 & $  0.57$  & $-33.22$  & $ -8.38$  & $ -0.13$  & $-59.94$  & $  5.15$ \\
422.8263 & 150 & $  2.78$  & $-33.02$  & $ -9.19$  & $ -1.85$  & $-62.53$  & $  6.09$ \\
423.6041 & 179 & $  0.07$  & $-49.87$  & $-14.35$  & $  4.62$  & $-85.41$  & $  3.43$ \\
423.6250 & 195 & $  0.07$  & $-51.39$  & $-16.90$  & $  1.32$  & $-89.97$  & $  5.21$ \\
423.6460 & 189 & $ -0.64$  & $-52.35$  & $-14.95$  & $  0.00$  & $-89.23$  & $  2.71$ \\
423.6669 & 151 & $  1.21$  & $-50.08$  & $-12.34$  & $ -1.39$  & $-88.43$  & $ -0.39$ \\
423.6879 & 192 & $ -6.78$  & $-52.29$  & $-14.01$  & $  3.04$  & $-91.02$  & $  4.43$ \\
423.7088 & 163 & $ -7.99$  & $-51.32$  & $ -5.83$  & $ -1.58$  & $-91.45$  & $  1.99$ \\
423.7297 & 159 & $ -5.50$  & $-46.83$  & $-10.26$  & $ -1.25$  & $-91.02$  & $  2.27$ \\
423.7507 & 173 & $ -1.21$  & $-45.79$  & $-12.27$  & $ -1.72$  & $-95.28$  & $  3.88$ \\
423.7716 & 161 & $  1.36$  & $-51.11$  & $-11.53$  & $ -2.64$  & $-94.29$  & $  5.48$ \\
423.7925 & 219 & $ -6.50$  & $-39.16$  & $-11.60$  & $ -4.62$  & $-95.90$  & $ -0.72$ \\
423.8135 & 309 & $ -6.85$  & $-44.62$  & $-11.46$  & $ -2.44$  & $-96.20$  & $  4.65$ \\
423.8344 & 241 & $ -5.42$  & $-50.77$  & $-14.28$  & $ -3.30$  & $-95.83$  & $  0.28$ \\
423.8553 & 185 & $ -5.35$  & $-48.90$  & $-12.94$  & $ -3.04$  & $-99.23$  & $  3.49$ \\
\hline
  \end{tabular}
  \end{center}
\end{table*}

\addtocounter{table}{-1}
\begin{table*}[h!]
  \caption{continued}
  \begin{center}
  \begin{tabular}{c c r r r r r r}
\hline
HJD-2459000 & S/N & \multicolumn{6}{c}{RV (km\,s$^{-1}$)} \\
& & He\,{\sc ii} $\lambda$\,4200 & H$\gamma$ & He\,{\sc i} $\lambda$\,4471 & He\,{\sc ii} $\lambda$\,4542 & H$\beta$ & He\,{\sc ii} $\lambda$\,5412 \\
\hline
423.8763 & 229 & $ -5.21$  & $-40.89$  & $-13.28$  & $ -2.84$  & $-99.84$  & $  6.92$ \\
425.7029 & 281 & $  0.79$  & $-26.59$  & $ -2.95$  & $ -2.31$  & $-38.05$  & $  3.55$ \\
425.7238 & 201 & $ -2.57$  & $-22.72$  & $ -0.34$  & $  2.84$  & $-34.84$  & $  1.00$ \\
425.7447 & 212 & $  0.64$  & $-20.51$  & $ -2.01$  & $ -2.77$  & $-33.92$  & $  9.14$ \\
425.7657 & 177 & $ -0.07$  & $-18.92$  & $ -6.17$  & $ -2.97$  & $-35.09$  & $  9.47$ \\
425.7866 & 162 & $ -5.07$  & $-25.76$  & $ -5.30$  & $  1.85$  & $-32.44$  & $ 10.97$ \\
425.8075 & 185 & $ -6.28$  & $-22.86$  & $ -3.75$  & $ -0.13$  & $-35.77$  & $ 10.64$ \\
425.8285 & 170 & $  0.57$  & $-25.00$  & $ -5.23$  & $ -0.79$  & $-43.97$  & $  2.83$ \\
425.8494 & 315 & $  2.78$  & $-26.73$  & $ -7.64$  & $ -0.53$  & $-34.04$  & $  2.77$ \\
425.8703 & 229 & $ -4.21$  & $-31.08$  & $ -4.22$  & $ -1.98$  & $-39.59$  & $  5.82$ \\
425.8913 & 198 & $ -3.00$  & $-32.26$  & $-12.00$  & $ -1.19$  & $-45.51$  & $  6.32$ \\
427.6479 & 197 & $ -2.43$  & $-36.40$  & $ -9.45$  & $  0.86$  & $-52.54$  & $  1.88$ \\
427.6689 & 145 & $ -2.93$  & $-34.95$  & $-12.87$  & $ -0.20$  & $-52.42$  & $  3.10$ \\
427.6898 & 185 & $ -7.28$  & $-33.02$  & $-12.54$  & $ -2.51$  & $-53.90$  & $ -1.22$ \\
427.7108 & 146 & $ -3.93$  & $-36.74$  & $-12.74$  & $ -2.44$  & $-55.93$  & $  1.33$ \\
427.7317 & 176 & $ -5.64$  & $-35.99$  & $-11.40$  & $ -1.52$  & $-54.39$  & $ -1.55$ \\
427.7824 & 205 & $ -5.35$  & $-35.43$  & $-13.61$  & $  0.20$  & $-55.81$  & $  2.16$ \\
427.8033 & 195 & $ -4.57$  & $-34.81$  & $-13.21$  & $  0.00$  & $-56.30$  & $  1.27$ \\
427.8243 & 197 & $ -2.78$  & $-35.29$  & $-13.21$  & $  0.59$  & $-54.95$  & $ -0.06$ \\
427.8452 & 249 & $ -0.86$  & $-33.98$  & $ -9.25$  & $  0.13$  & $-55.56$  & $  3.49$ \\
427.8661 & 232 & $  0.71$  & $-32.60$  & $ -9.12$  & $  2.90$  & $-52.48$  & $  5.65$ \\
427.8871 & 170 & $  2.14$  & $-30.67$  & $ -8.92$  & $  4.36$  & $-52.30$  & $  8.70$ \\
427.9080 & 242 & $  3.50$  & $-27.63$  & $ -9.05$  & $  0.92$  & $-51.62$  & $  9.97$ \\
428.7254 & 187 & $-10.64$  & $-33.64$  & $-13.74$  & $ -3.43$  & $-49.52$  & $ -2.38$ \\
428.7464 & 203 & $  2.78$  & $-33.84$  & $-15.76$  & $ -5.94$  & $-51.43$  & $ -7.92$ \\
428.7673 & 215 & $ -8.42$  & $-29.29$  & $-16.16$  & $ -3.83$  & $-48.78$  & $ -2.16$ \\
428.7882 & 202 & $ -8.99$  & $-34.05$  & $-15.09$  & $ -5.61$  & $-48.47$  & $ -4.27$ \\
428.8092 & 232 & $ -8.92$  & $-33.15$  & $-12.94$  & $ -2.38$  & $-51.43$  & $ -0.55$ \\
428.8301 & 153 & $ -6.42$  & $-32.05$  & $-12.74$  & $ -6.60$  & $-50.32$  & $ -4.60$ \\
437.7529 & 181 & $ -0.07$  & $-36.95$  & $-12.14$  & $ -2.84$  & $-58.34$  & $  4.15$ \\
438.6223 & 198 & $  2.93$  & $-18.10$  & $  0.47$  & $  7.46$  & $-32.31$  & $  6.37$ \\
438.6433 & 203 & $  1.00$  & $-19.06$  & $  0.47$  & $  5.02$  & $-35.34$  & $  9.20$ \\
438.6642 & 167 & $  0.21$  & $-20.44$  & $ -0.13$  & $  4.82$  & $-32.56$  & $  9.75$ \\
438.6851 & 265 & $  3.85$  & $-19.62$  & $ -0.40$  & $  3.04$  & $-35.89$  & $  4.38$ \\
438.7061 & 204 & $  1.86$  & $-23.14$  & $ -1.48$  & $  1.19$  & $-34.97$  & $  6.98$ \\
438.7270 & 167 & $ -2.07$  & $-18.86$  & $ -1.41$  & $  1.98$  & $-34.90$  & $  6.37$ \\
438.7480 & 224 & $ -4.64$  & $-24.31$  & $ -4.22$  & $  1.52$  & $-36.82$  & $  3.66$ \\
438.7689 & 237 & $ -4.14$  & $-25.42$  & $ -3.82$  & $ -4.75$  & $-42.18$  & $  1.38$ \\
438.7898 & 206 & $ -5.92$  & $-26.32$  & $ -9.59$  & $ -3.76$  & $-40.02$  & $  1.55$ \\
438.8108 & 225 & $ -2.86$  & $-28.80$  & $ -7.44$  & $ -2.97$  & $-44.83$  & $  1.38$ \\
442.6029 & 201 & $  2.93$  & $-26.87$  & $ -2.48$  & $  3.89$  & $-49.58$  & $ 10.47$ \\
442.6239 & 212 & $  1.64$  & $-26.18$  & $ -2.21$  & $  4.49$  & $-48.35$  & $  9.69$ \\
442.6448 & 163 & $  4.00$  & $-26.25$  & $ -1.54$  & $  0.26$  & $-47.61$  & $  8.92$ \\
442.6658 & 195 & $  2.57$  & $-27.28$  & $ -4.22$  & $  8.45$  & $-46.62$  & $  6.92$ \\
442.7134 & 138 & $  1.78$  & $-26.59$  & $ -3.22$  & $  6.93$  & $-45.02$  & $  9.58$ \\
442.7645 & 125 & $  3.07$  & $-25.69$  & $ -2.82$  & $  2.64$  & $-46.37$  & $  6.37$ \\
446.6729 & 125 & $ -2.00$  & $-28.66$  & $ -8.78$  & $  1.58$  & $-46.87$  & $  5.37$ \\
450.6779 & 247 & $ -4.43$  & $-34.95$  & $-17.97$  & $ -4.82$  & $-57.48$  & $  1.83$ \\
453.6701 & 235 & $ -0.29$  & $-24.11$  & $ -3.89$  & $ -1.12$  & $-32.13$  & $  4.99$ \\
499.5658 & 220 & $ -6.28$  & $-34.81$  & $-14.88$  & $ -2.90$  & $-53.28$  & $ -3.21$ \\
\hline
  \end{tabular}
  \end{center}
  \tablefoot{The first and second columns yield the heliocentric julian date at mid-exposure and the signal-to-noise ratio evaluated in the wavelength range between 5880\,\AA\ and  5886\,\AA. The subsequent columns indicate the heliocentric RVs measured on prominent absorption lines. The adopted rest wavelengths were 4199.87\,\AA, 4340.47\,\AA, 4471.48\,\AA, 4541.59\,\AA, 4861.33\,\AA, and 5411.52\,\AA.}
\end{table*}
  \end{appendix}
\end{document}